\documentclass[pre,twocolumn,amsmath,showpacs,superscriptaddress]{revtex4}

\usepackage{amsmath}
\usepackage{amsbsy}
\usepackage{amscd}
\usepackage{amsopn}
\usepackage{amstext}
\usepackage{amsxtra}
\usepackage{epsfig}
\usepackage{rotating}

\def\figs{figs/}
\def\sig{\sigma}
\def\WT{\widetilde}
\def\lb{\left}
\def\rb{\right}
\def\D{\Delta}
\def\lam{\lambda}
\def\g{\gamma}

\begin{document}

\title{The Unreasonable Effectiveness of Tree-Based Theory for Networks with Clustering}

\author{Sergey Melnik}
\affiliation{Department of Mathematics \& Statistics, University of Limerick, Ireland}
\author{Adam Hackett}
\affiliation{Department of Mathematics \& Statistics, University of Limerick, Ireland}
\author{Mason A. Porter}
\affiliation{Oxford Centre for Industrial and Applied Mathematics, Mathematical Institute, University of Oxford, OX1 3LB, UK}
\affiliation{CABDyN Complexity Centre, University of Oxford, Oxford OX1 1HP, UK}
\author{Peter J. Mucha}
\affiliation{Carolina Center for Interdisciplinary Applied Mathematics, Department of Mathematics, University of North Carolina, Chapel Hill, NC 27599-3250, USA}
\affiliation{Institute for Advanced Materials, Nanoscience \& Technology, University of North Carolina, Chapel Hill, NC 27599-3216, USA}
\author{James P. Gleeson}
\affiliation{Department of Mathematics \& Statistics, University of Limerick, Ireland}

\begin{abstract}
We demonstrate that a tree-based theory for various dynamical processes yields extremely accurate results for several networks with high levels of clustering. We find that such a theory works well as long as the mean intervertex distance $\ell$ is sufficiently small---i.e., as long as it is close to the value of $\ell$ in a random network with negligible clustering and the same degree-degree correlations. We confirm this hypothesis numerically using real-world networks from various domains and on several classes of synthetic clustered networks. We present analytical calculations that further support our claim that tree-based theories can be accurate for clustered networks provided that the networks are ``sufficiently small'' worlds.
\end{abstract}
\maketitle

\section{Introduction} \label{sec1}
One of the most important areas of network science is the study of dynamical processes on networks~\cite{Strogatz01,Boccaletti06,Barrat08,Dorogovtsev08}. On one hand, research on this topic has provided interesting theoretical challenges for physicists, mathematicians, and computer scientists. On the other hand, there is an increasing recognition of the need to improve the understanding of dynamical systems on networks to achieve advances in epidemic dynamics~\cite{Pastor-Satorras01,Barthelemy05,Eames08}, traffic flow in both online and offline systems~\cite{Traffic06}, oscillator synchronization~\cite{arenas08}, and more~\cite{Barrat08}.

Analytical results for complex networks are rather rare, especially if one wants to study a dynamical system on a network topology that attempts to incorporate even minimal features of real-world networks. If one considers a dynamical system on a real-world network rather than on a grossly simplified caricature of it, then theoretical results become almost barren. Furthermore, most analyses assume that the network under study has a locally tree-like structure, so that they can only possess very few small cycles, whereas most real networks have significant clustering (and, in particular, possess numerous small cycles). This has motivated a wealth of recent research concerning analytical results on networks with clustering~\cite{Newman03b,comnotices,Newman09,Gleeson09a,Gleeson09b,Ostilli09,Serrano06a,Serrano06b,Serrano06c,Trapman07,Miller09a,Miller09b,Eames08,Britton08}.

Most existing theoretical results for (unweighted) networks are derived for an ensemble of networks using (i) only their degree distribution $p_k$, which gives the probability that a random node has degree $k$ (i.e., has exactly $k$ neighbors) or using (ii) their degree distribution and their degree-degree correlations, which are defined by the joint degree distribution $P(k,k')$ describing the probability that a random edge joins nodes of degree $k$ and $k'$. In the rest of this paper, we will refer to case (i) as ``$p_k$-theory'' (the associated random graph ensemble is known as the ``configuration model'' \cite{Newman03a}) and to case (ii) as ``$P(k,k')$-theory''. The clustering in sample networks is low in both situations; it typically decreases as $N^{-1}$ as the number of nodes $N \rightarrow \infty$~\footnote{We assume that the degree distribution has finite variance, as real-world networks necessarily have a finite cutoff in their degree sequence.}.

We concentrate in this paper on undirected, unweighted real-world networks, which can be described completely using adjacency matrices. It is straightforward to calculate the empirical distributions $p_k$ and $P(k,k')$, which can then be used as inputs to analytical theory for various well-studied processes. The results can subsequently be compared with large-scale numerical simulations using the original networks.
\begin{figure*}
\centering
\includegraphics[width=1.95\columnwidth]{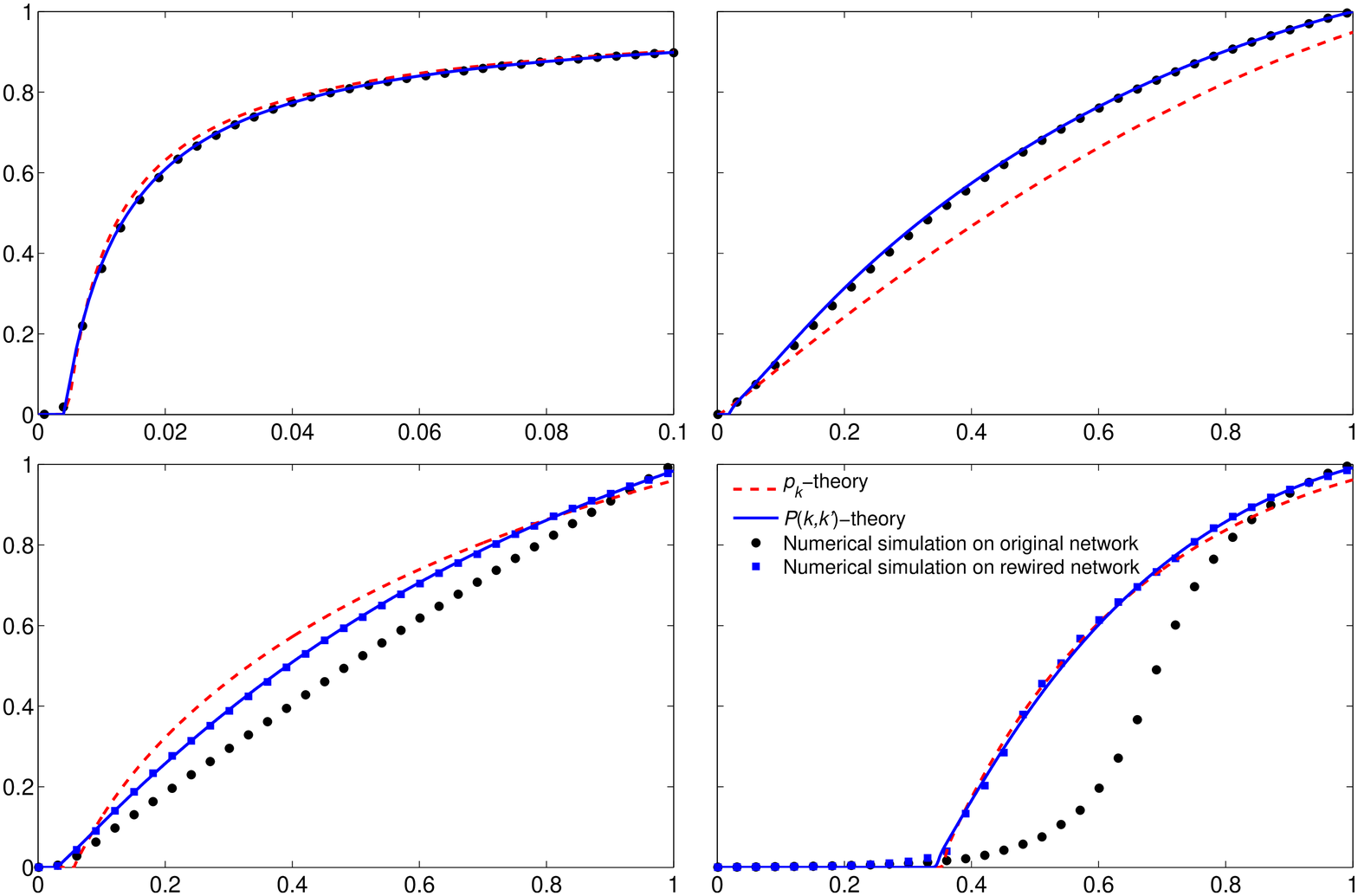}
\put(-380,180){\bf (a) Facebook Oklahoma}
\put(-80,180){\bf (b) AS Internet}
\put(-380,20){\bf (c) PGP Network}
\put(-80,20){\bf (d) Power Grid}
\put(-240,-15){\bf \begin{LARGE}$p$\end{LARGE}}
\put(-495,150){\bf \begin{LARGE}$S$\end{LARGE}}
\caption{(Color online) Bond percolation. Plots of GCC size $S$ versus bond occupation probability $p$ for various real-world networks. These networks, which we also use as examples in other figures, are (a) the Facebook network for University of Oklahoma~\cite{Traud08}, (b) the Internet at the AS level~\cite{asrel_url}, (c) the PGP network~\cite{Guardiola02, Boguna04, PGP_url}, and (d) the power grid for the western United States~\cite{Watts98, power_url};
}
\label{fig1}
\end{figure*}

In the present paper, we demonstrate that analytical results derived using tree-based theory can be applied with high accuracy to certain networks despite their high levels of clustering. Examples of such networks include university social networks constructed using Facebook data~\cite{Traud08} and the Autonomous Systems (AS) Internet graph~\cite{asrel_url}. Specifically, the analytical results for bond percolation, $k$-core sizes, and other processes accurately match simulations on a given (clustered) network provided that the mean intervertex distance in the network is sufficiently small---i.e., that it is close to its value in a randomly rewired version of the graph. Recalling that a clustered network with a low mean intervertex distance is said to have the \emph{small-world property}, we find that tree-based analytical results are accurate for networks that are ``sufficiently small'' small worlds. In discussing this result, we focus considerable attention on quantifying what it means to be ``sufficiently small''.

The remainder of this paper is organized as follows. In Sec.~\ref{sec2}, we consider several dynamical processes on highly clustered networks and show that tree-based theory adequately describes them on certain networks but not on others. In order to explain our observations, we introduce in Sec.~\ref{sec3} a measure of prediction quality $E$ and develop a hypothesis, inspired by the well-known Watts-Strogatz example of small-world networks, regarding its dependence on the mean intervertex distance $\ell$. We provide support for our hypothesis by numerical examination of a large range of networks in Appendix~\ref{appB} and by analytical calculations in Appendix~\ref{appA}. We discuss our conclusions in Sec.~\ref{sec5}.

\section{Dynamical Processes on Networks} \label{sec2}
\subsection{Bond Percolation}

We begin by considering bond percolation, which has been studied extensively on networks. In bond percolation, network edges are deleted (or labeled as \emph{unoccupied}) with probability $1- p$, where $p$ is called the \emph{bond occupation probability}. One can measure the effect of such deletions on the aggregate graph connectivity in the limit of infinitely many nodes using $S(p)$, the fractional size of the giant connected component (GCC) at a given value of $p$. (In this paper, we will use the terminology GCC for finite graphs as well.) Bond percolation has been used in simple models for epidemiology. In such a context, $p$ is related to the average transmissibility of a disease, so that the GCC is used to represent the size of an epidemic outbreak (and to give the steady-state infected fraction in an susceptible-infected-recovered model)~\cite{Newman03a}.
\begin{figure*}
\centering
\includegraphics[width=1.95\columnwidth]{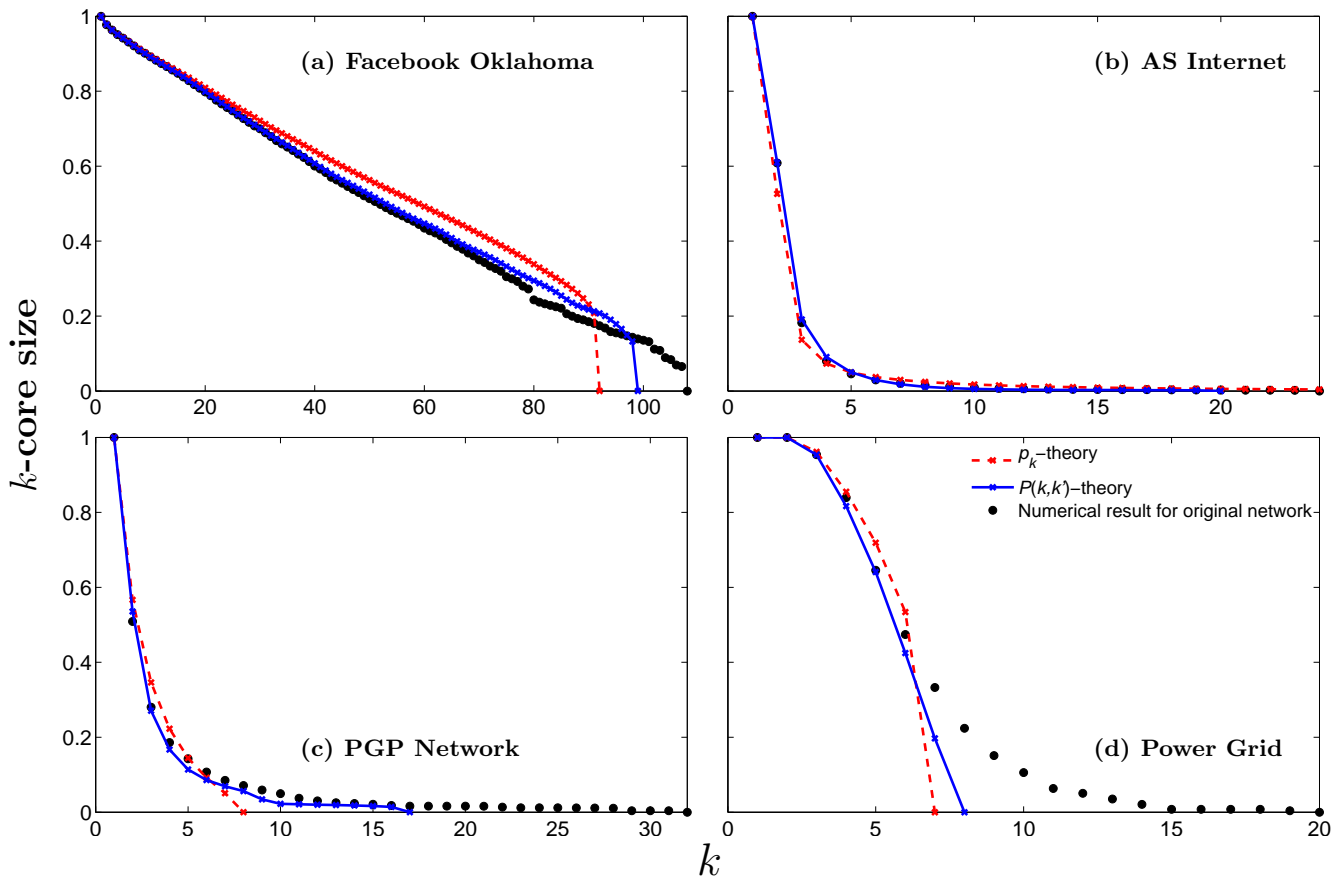}
\put(-390,290){\bf (a) Facebook Oklahoma}
\put(-90,290){\bf (b) AS Internet}
\put(-390,30){\bf (c) PGP Network}
\put(-90,30){\bf (d) Power Grid}
\put(-240,-15){\bf \begin{LARGE}$k$\end{LARGE}}
\put(-490,130){\begin{rotate}{90}\bf \begin{Large}$k$-core size\end{Large}\end{rotate}}
\caption{(Color online) Plots of $k$-core sizes versus $k$ for the real-world networks from Fig.~\ref{fig1}. The highest nonzero $k$-cores are
(a) $K_{p_k}=91$, $K_{P(k,k')}=98$, $K_{\rm num}=107$;
(b) $K_{p_k}=132$, $K_{P(k,k')}=19$, $K_{\rm num}=23$;
(c) $K_{p_k}=7$, $K_{P(k,k')}=16$, $K_{\rm num}=31$; and
(d) $K_{p_k}=6$, $K_{P(k,k')}=7$, $K_{\rm num}=19$.
}
\label{fig2}
\end{figure*}

Analytical results for GCC sizes for $p_k$-theory \cite{Callaway00} can be found in Eq.~(8.11) of Ref.~\cite{Newman03a} and analytical results for $P(k,k')$-theory are available in Eq.~(12) of Ref.~\cite{Vazquez03}. We plot these theoretical predictions in Fig.~\ref{fig1} as dashed red and solid blue curves, respectively. In this figure, we use the following data sets as examples: (a) the September 2005 Facebook network for University of Oklahoma~\cite{Traud08}, where nodes are people and links are friendships; (b) the Internet at the Autonomous Systems (AS) level~\cite{asrel_url}, where nodes represent ASs and links indicate the presence of a relationship; (c) the network of users of the Pretty-Good-Privacy (PGP) algorithm for secure information interchange~\cite{Guardiola02, Boguna04, PGP_url}; and (d) the network representing the topology of the power grid of the western United States~\cite{Watts98, power_url}. We treat all data sets as undirected, unweighted networks.

We performed numerical calculations of the GCC size using the algorithm in Ref.~\cite{Newman01e} and plotted the results as black disks in Fig.~\ref{fig1}. It is apparent from Fig.~\ref{fig1}(a,b) that $P(k,k')$-theory matches numerical simulations very accurately for the AS Internet and Oklahoma Facebook networks, and we found similar accuracy for all 100 single-university Facebook data sets available to us. However, as shown in Fig.~\ref{fig1}(c,d), the match between theory and numerics is much poorer on the PGP and Power Grid networks. The usual explanation for this lack of accuracy is that it is caused by clustering in the real-world network that is not captured by $P(k,k')$-theory. Note, however, that the Oklahoma Facebook network has one of the highest clustering coefficients of the four cases in Fig.~\ref{fig1} even though it is accurately described by its $P(k,k')$-theory.

Indeed, the global clustering coefficients (defined as the mean of the local clustering coefficient over all nodes~\cite{Watts98}) for the Oklahoma Facebook, AS Internet, PGP, and Power Grid networks are 0.23, 0.21, 0.27, and 0.08, respectively. (See Table~\ref{table1} for basic summary statistics for these networks.) The clustering coefficients for all 100 Facebook networks range from 0.19 to 0.41, and the mean value of these coefficients is 0.24. These observations suggest that one ought to consider other explanatory mechanisms for the discrepancy between theory and simulations in Fig.~\ref{fig1}(c,d).
\begin{figure*}
\centering
\includegraphics[width=1.95\columnwidth]{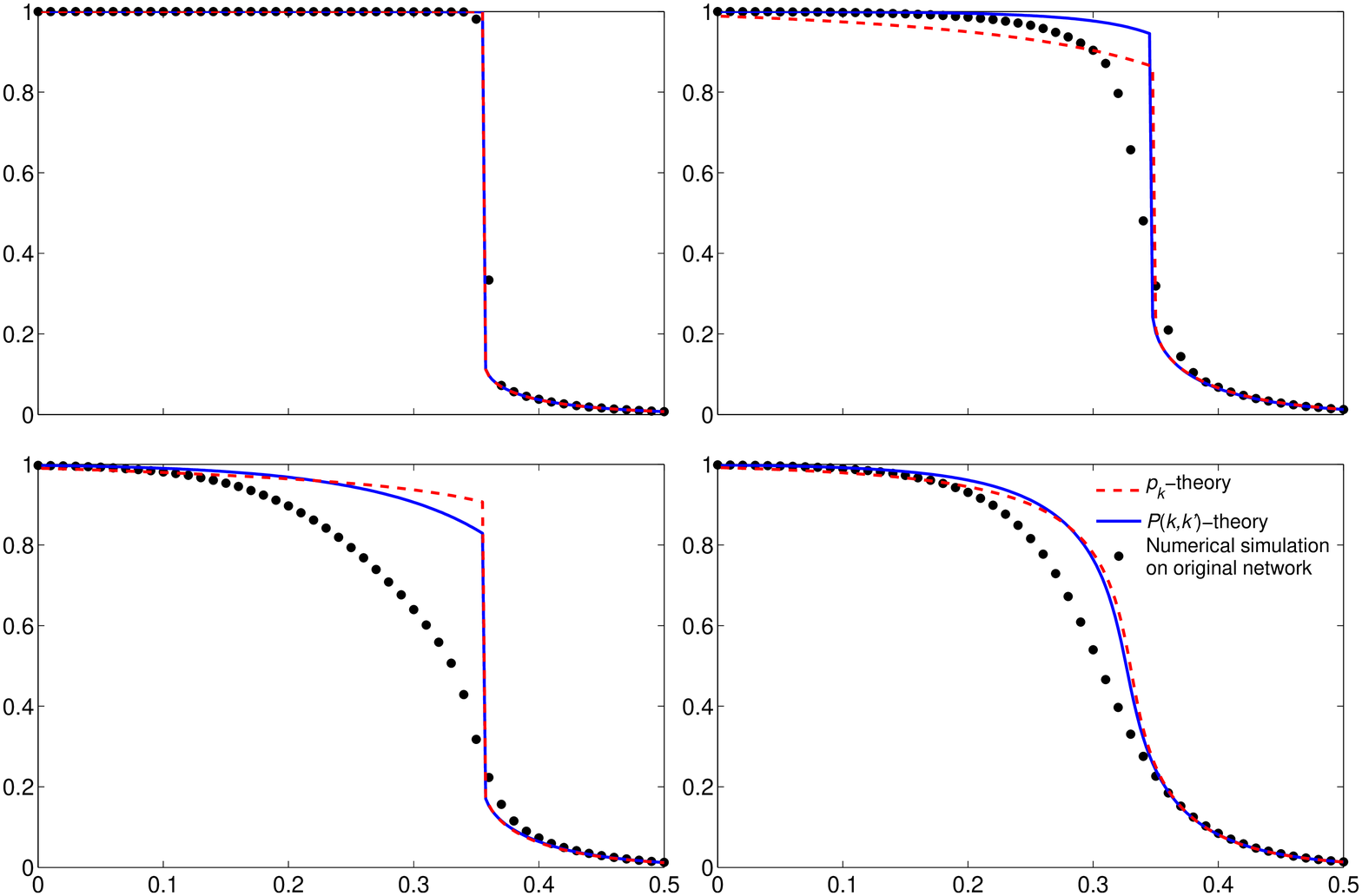}
\put(-450,180){\bf (a) Facebook Oklahoma} \put(-200,180){\bf
(b) AS Internet} \put(-450,20){\bf (c) PGP Network} \put(-200,20){\bf (d)
Power Grid} \put(-240,-15){\bf \begin{LARGE}$\mu$\end{LARGE}}
\put(-490,150){\bf \begin{LARGE}$\rho$\end{LARGE}} 
\caption{(Color online) Watts' threshold model, with threshold mean $\mu$ and variance $\sig^2 = 0.04$ for the networks from Fig.~\ref{fig1}. We use the seed fraction $\rho_0=0$ because the nodes with negative thresholds immediately turn on and act as initial seeds. In other words, the effective seed fraction is given by the cumulative distribution of thresholds at zero: $\lb[1+ {\rm erf}\lb(-\mu/\lb(\sig\sqrt{2}\rb)\rb)\rb]/2$.}
\label{fig3}
\end{figure*}

In considering other explanations, note that the discrepancy between theory and numerics in Fig.~\ref{fig1}(c,d) does not arise from finite-size effects. To demonstrate this, we rewired the networks using an algorithm that preserves the $P(k,k')$ distribution but otherwise randomizes connections between the $N$ nodes~\footnote{We employ the following network rewiring algorithm: Choose an edge of the network at random. Denote its associated vertices by $A$ and $B$ and their corresponding degrees by $k_A$ and $k_B$. From the set of edges that are connected to one vertex of degree $k_A$, choose another edge at random. This edge connects the vertices $C$ and $D$, whose respective degrees are $k_A$ and $k_D$. Now rewire the two chosen edges to obtain the edges $AD$ and $CB$ instead of $AB$ and $CD$. This rewiring scheme does not affect the degrees of the rewired vertices, but applying it repeatedly significantly reduces the local clustering (i.e., the density of triangles). In applying this algorithm, we also take care to avoid multiple and self-links.}. Because this scheme preserves the degree correlation matrix $P(k,k')$, we call this the \emph{$P$-rewiring} algorithm. Note that the ensemble of fully $P$-rewired networks is in fact the ensemble of random networks defined by the $P(k,k')$ matrix of the original (unrewired) network.

We show numerical calculations of the GCC sizes for these rewired networks with blue squares in Fig.~\ref{fig1}(c,d) and observe that they agree very well with the curves produced from $P(k,k')$-theory. We conclude that the structural characteristics of the original networks---rather than simply their sizes---must underlie the observed differences between simulations and analytics.

Also note that the agreement between $P(k,k')$- and $p_k$-theories in Fig.~\ref{fig1} is better in panels (a) and (d) than in panels (b) and (c). This is because the Pearson correlation coefficient $r$ of the end-vertex degrees of a random edge~\cite{Newman03a} has smaller absolute values for the networks shown in panels (a) and (d) (0.074, with the mean 0.063 over 100 Facebook networks, and 0.0035, respectively) than it does for the networks in (b) and (c) ($-0.2$ and 0.24, respectively).

\subsection{$k$-Cores}
\begin{figure*}
\centering
\includegraphics[width=1.95\columnwidth]{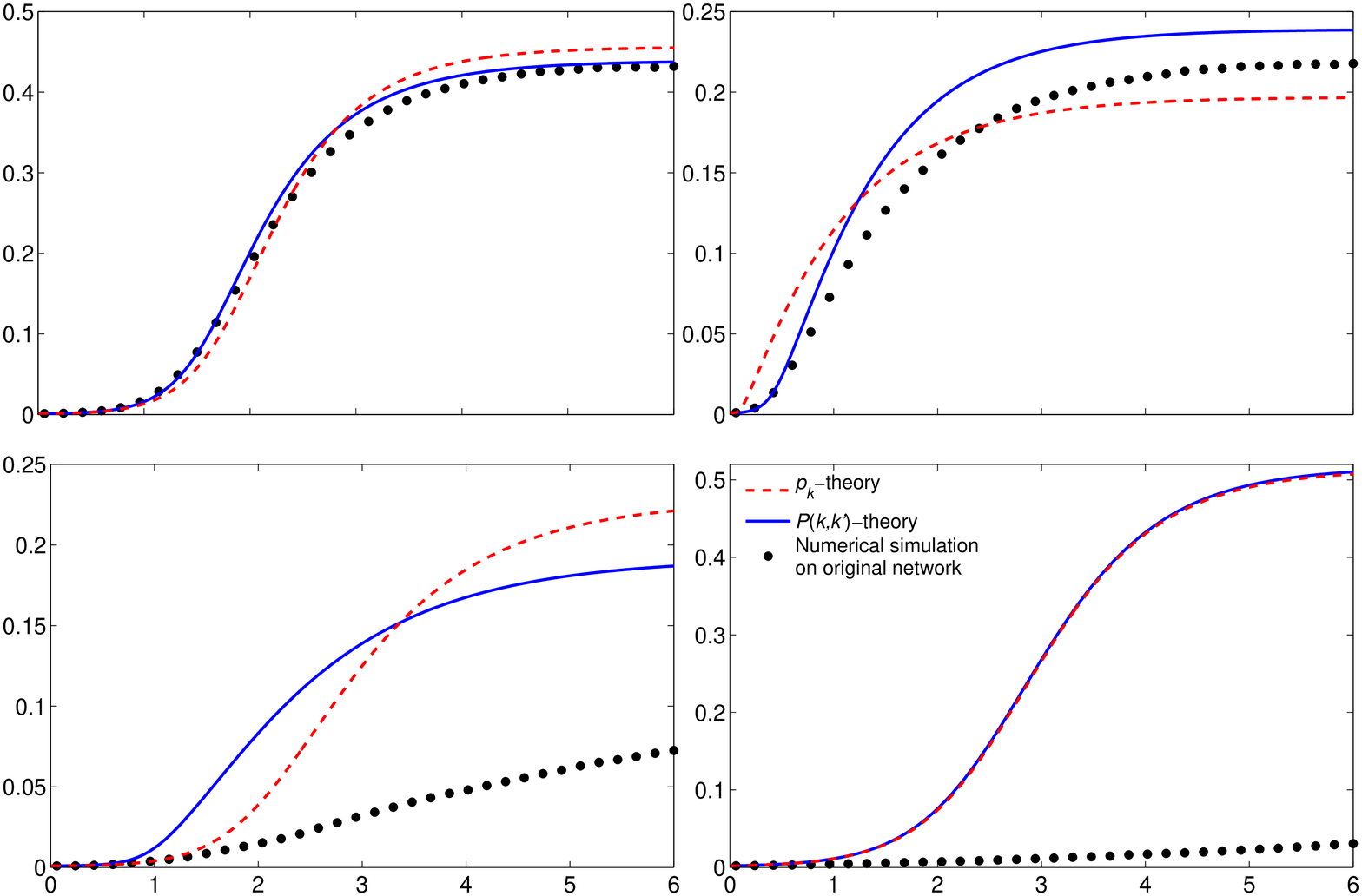}
\put(-390,180){\bf (a) Facebook Oklahoma} \put(-90,180){\bf
(b) AS Internet} \put(-330,20){\bf (c) PGP Network} \put(-90,20){\bf (d)
Power Grid} \put(-240,-15){\bf \begin{LARGE}$t$\end{LARGE}}
\put(-510,150){\bf \begin{LARGE}$I(t)$\end{LARGE}} 
\caption{(Color online) SIS dynamics, which we display as plots of infected fraction $I(t)$ versus time $t$ for the networks from Fig.~\ref{fig1}. The parameters in Eq.~(17) of Ref.~\cite{Barthelemy05} are the recovery rate $\mu$ and the spreading rate $\lam$. We use the value $\mu = 1$ in all figure panels; we use $I(0) = 10^{-3}$ in panels (a)--(c) and $I(0)=0.002$ in panel (d); and we use $\lam=0.02$ in panel (a); $\lam=0.2$ in (b) and (c); and $\lam=0.8$ in panel (d).}
\label{fig4}
\end{figure*}

Figures~\ref{fig2},~\ref{fig3}, and~\ref{fig4} show similar comparisons of analytical results versus numerical simulations for other well-studied processes on networks. In Fig.~\ref{fig2}, we plot the $k$-core sizes of the networks. The $k$-core is the largest subgraph whose nodes all have degree at least $k$. The $p_k$-theory for $k$-core sizes is given in Ref.~\cite{Dorogovtsev06} and the $P(k,k')$-theory is given by Eq.~(32) of Ref.~\cite{Gleeson08a}. As shown in Fig.~\ref{fig2}(a,b), we again find very good agreement of $P(k,k')$-theory with numerical calculations on the AS Internet and Facebook networks and less accurate results for the other example networks. This can be quantified by comparing the actual (numerical) result for the highest value of $k$ for which the $k$-core size is nonzero to the value that is predicted by $P(k,k')$-theory. (We use $K$ to denote this maximal value of $k$.) For Fig.~\ref{fig2}(a) and (b), we obtain $K_{P(k,k')}/K_{\rm num} \approx 0.916$ and $K_{P(k,k')}/K_{\rm num} \approx 0.826$, respectively. The corresponding values for Fig.~\ref{fig2}(c) and (d) are $K_{P(k,k')}/K_{\rm num} \approx 0.516$ and $K_{P(k,k')}/K_{\rm num} \approx 0.368$.

\subsection{Watts' Threshold Model}
Watts~\cite{Watts02} introduced a simple model for the spread of cultural fads. It allows one to examine how a small initial fraction of early adopters can lead to a global cascade of adoption via a social network. The $p_k$-theory and $P(k,k')$-theory for the average cascade size are given, respectively, in Ref.~\cite{Gleeson07a} and Ref.~\cite{Gleeson08a}. In Fig.~\ref{fig3}, we compare these theories with numerical simulations on populations with Gaussian threshold distributions of mean $\mu$ and variance $\sig^2 = 0.04$. The cascade size shows a sharp transition as $\mu$ is increased. As with the other processes that we discussed above, the position of this transition is accurately captured by the theory for the Facebook and AS Internet networks but not for the other examples. 

\subsection{Susceptible-Infected-Susceptible Model}
In Fig.~\ref{fig4}, we show a comparison between theory and numerical simulation results for the time evolution of a susceptible-infected-susceptible (SIS) epidemic model on various networks. Unlike the other processes that we have discussed, the theory for this case---as given, for example, by Eq.~(17) of Ref.~\cite{Barthelemy05}---is expected to apply accurately only to the early-time development of the infection. In view of this restriction, the results of Fig.~\ref{fig4} are consistent with those of Figs.~\ref{fig1}--\ref{fig3}. That is, the $P(k,k')$-theory once again provides accurate results for certain networks for a variety of processes of interest but is rather inaccurate for other networks.

\section{Measure of Prediction Quality} \label{sec3}
We now aim to characterize the types of networks for which $P(k,k')$-theory can be expected to give good results. Because Figs.~\ref{fig1}--\ref{fig4} demonstrate that this characterization holds for several processes, we will concentrate hereafter primarily on the bond percolation case.

\subsection{Watts-Strogatz Networks}
\begin{figure*}[floatfix]
\centering
\includegraphics[width=1.3\columnwidth]{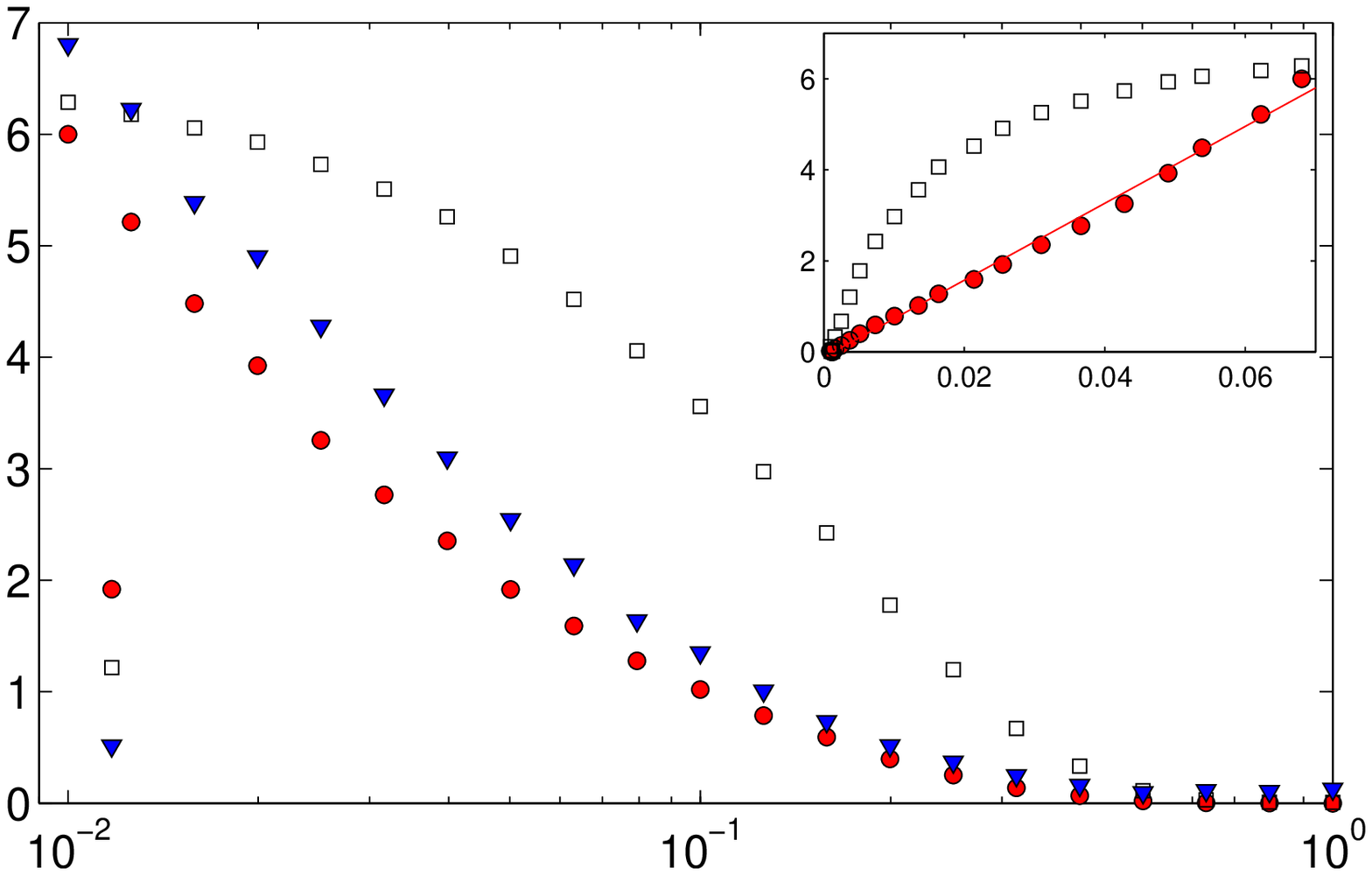}
\put(-160,-15){\bf \begin{large}$f$\end{large}} 
\put(-360,120){\bf\begin{large}$\ell_f-\ell_1$\end{large}} 
\put(-365,100){\bf\begin{large}$10\times C_f$\end{large}} 
\put(-370,80){\bf\begin{large}$100\times E_f$\end{large}} 
\put(-80,105){\bf\begin{small}$E_f$\end{small}} 
\put(-165,162){\bf\begin{small}$\ell_f-\ell_1$\end{small}} 
\put(-165,150){\bf\begin{small}$10\times C_f$\end{small}} 
\put(-285,65){\bf\begin{normalsize}$\ell_f-\ell_1$\end{normalsize}} 
\put(-285,45){\bf\begin{normalsize}$C_f$\end{normalsize}} 
\put(-285,25){\bf\begin{normalsize}$E_f$\end{normalsize}} 
\caption{(Color online) Watts-Strogatz small-world network: $\ell_f-\ell_1$ (red circles), $10\times C_f$ (open squares), and $100\times E_f$ (blue triangles) as functions of rewiring fraction $f$. The inset shows $\ell_f - \ell_1$ and $C_f$ as functions of $E_f$ for $f\ge 10^{-2}$. Observe the linear relation between $E_f$ and $\ell_f-\ell_1$, which suggests that $\ell_f-\ell_1$ might be a good indicator of how well the bond-percolation process on a network can be approximated by tree-based theory.} 
\label{fig5}
\end{figure*}

Using the small-world networks introduced by Watts and Strogatz \cite{Watts98}, one can conduct a systematic study of the effects of clustering $C$ and the mean intervertex distance $\ell$. We start with a ring of $N = 10000$ nodes and connect each node to $z=10$ nearest neighbors. We then randomly rewire a fraction $f$ of the links in the network~\footnote{We employ our $P$-rewiring algorithm that preserves the degree of each node, which is slightly different from the one used in~\cite{Watts98}, but this difference is not important for the phenomenon under study.}. When $f=0$, the values of $C$ and $\ell$ are both high. When $f=1$, the rewired network is connected completely at random, which gives it low $C$ and $\ell$ values. For each value of $f$ between 0 and 1, we numerically calculate the clustering coefficient $C_f$, the mean intervertex distance $\ell_f$, and the GCC size $S_f(p)$ for all values of the bond occupation probability $p$ between 0 and 1. The difference between $S_f(p)$ and the $P(k,k')$-theory curve, which we denote by $S_{\rm th}(p)$, gives a quantitative measure for the inaccuracy of the theory for this particular value of the rewiring parameter $f$. We define the error measure
\begin{equation}
E_f = \frac{1}{M} \sum_{i=1}^M \lb| S_{\rm th}(p_i)-S_f(p_i)\rb|\,, \label{e1}
\end{equation}
where $p_i=i/M$ for $i=1,2,\ldots,M$ are uniformly-spaced values in the interval $[0,1]$. Taking the spacing $1/M$ to be sufficiently fine (we use $1/M=10^{-3}$) implies that the error measure $E_f$ approaches the average vertical distance between the $S_{\rm th}(p)$ and $S_f(p)$ curves for $p \in [0,1]$.

In Fig.~\ref{fig5}, we plot the values of $\ell_f-\ell_1$, $C_f$ (scaled by a factor of 10 for ease of visualization), and $E_f$ (scaled by a factor of 100) as functions of the rewiring parameter $f$. For values of $f$ greater than $10^{-2}$, the quantities $\ell_f$ and $E_f$ exhibit similar behavior, whereas $C_f$ remains near its $f=0$ value of $2/3$ until $f$ is much larger~\footnote{When $f \ll 10^{-2}$, the quantity $\ell_f$ changes much more rapidly with $f$ than $E_f$ does. We focus on the range $f\ge 10^{-2}$ in Fig.~\ref{fig5} because for lower $f$, the values of the error $E_f$ are much larger than those seen in any of the networks we study (e.g., the Power Grid network has $E \approx 0.11$ and the PGP network has $E \approx 0.065$, which should be compared to the maximum error of $0.07$ seen in Fig.~\ref{fig5}).}. We highlight the similar scaling of $\ell_f$ and $E_f$ in the inset of Fig.~\ref{fig5}, in which we plot $\ell_f-\ell_1$ directly as a function of $E_f$ for $f\ge 10^{-2}$. The approximately linear dependence that we observe contrasts to the clearly nonlinear relation between $E_f$ and the clustering $C_f$ that we show in the same inset. This strongly suggests that differences between theory and numerics are related more directly to the mean intervertex distance than to the clustering coefficient.

\begin{figure*}
\centering
\includegraphics[width=0.95\columnwidth]{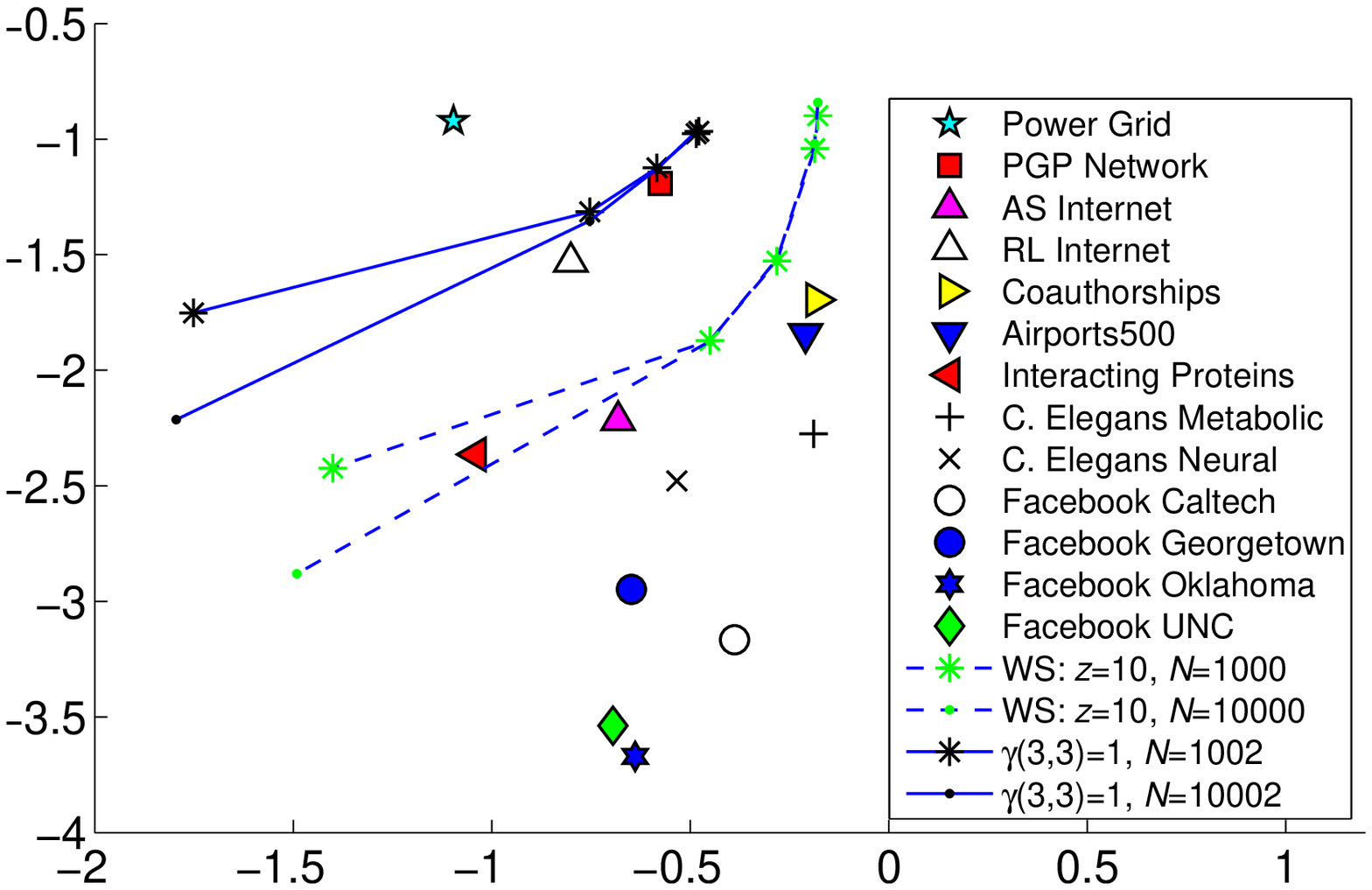}
\put(-210,140){\bf(a)}
\put(-140,-15){\bf \begin{large}$\log_{10} C$\end{large}}
\put(-240,65){\begin{rotate}{90} \bf\begin{large}$\log_{10} E$\end{large} \end{rotate}}
\hspace{1cm}
\includegraphics[width=0.95\columnwidth]{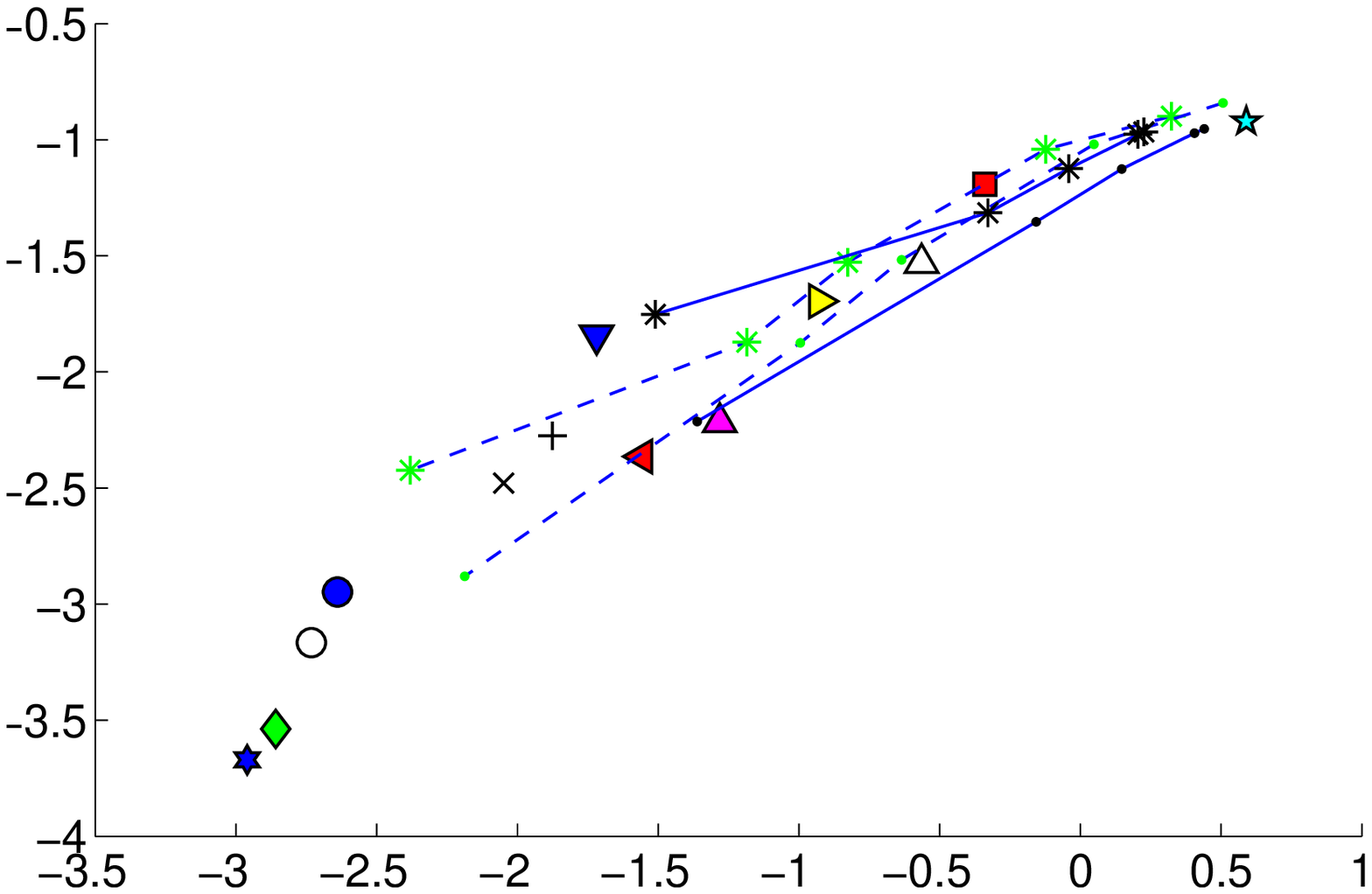}
\put(-210,140){\bf(b)} 
\put(-150,-15){\bf \begin{large}$\log_{10}[(\ell - \ell_{1})/z]$\end{large}} 
\put(-240,65){\begin{rotate}{90} \bf\begin{large}$\log_{10} E$\end{large} \end{rotate}}
\caption{(Color online) Scatter plots of $\log_{10} E$ versus (a) $\log_{10} C$ (with $R^2 \approx 0.087$) and (b) $\log_{10} \lb[(\ell-\ell_{1})/z\rb]$ (with $R^2 \approx 0.94$).}
\label{fig6}
\end{figure*}

\subsection{Real-World Networks and Additional Examples} \label{sect3B}
The above results for Watts-Strogatz small-world networks motivate the examination of a range of real-world networks in order to seek a clear relationship between an error measure similar to (\ref{e1}) and some other characteristic of the network, such as clustering or mean intervertex distance. For each network, we calculate the inaccuracy of $P(k,k')$-theory in terms of the error $E$, which measures the distance between the actual (numerically calculated) GCC size curve $S_{\rm num}(p)$ and the theoretical prediction $S_{\rm th}(p)$: 
\begin{equation}
E = \frac{1}{M} \sum_{i=1}^M \lb|
S_{\rm th}(p_i)-S_{\rm num}(p_i)\rb|\,.\label{e2}
\end{equation}
Essentially, $E$ gives the average distance between the numerics (black disks) and theory (solid blue curve) in Fig.~\ref{fig1}. In Fig.~\ref{fig6}(a), we show a scatter plot of $\log_{10}E$ versus $\log_{10}C$, where $C$ is the clustering coefficient of each network. We use logarithmic coordinates in Fig.~\ref{fig6} in order to fully resolve the range of values for both variables, as they vary by one or more orders of magnitude.

We also include synthetic examples, such as Watts-Strogatz small-world networks and clustered random networks generated using the recent models described in Refs.~\cite{Gleeson09a,Newman09}, which we now briefly recall. The fundamental quantity defining the $\g$-theory networks of Ref.~\cite{Gleeson09a} is the joint probability distribution $\g (k,c)$, which gives the probability that a randomly chosen node has degree $k$ and is a member of a $c$-clique (an all-to-all connected subgraph of $c$ nodes). With $\g(3,3)=1$ (and zero for other values of $k$ and $c$), each node in such a network has degree 3 and is part of exactly one triangle. This is equivalent to the $p_{1,1}=1$ case in the clustered random graph model of Ref.~\cite{Newman09}, where $p_{s,t}$ is the probability that a randomly chosen node is part of $t$ different triangles and in addition has $s$ single edges (which don't belong to the triangles). In each synthetic network, we $P$-rewire a fraction $f$ of links and show our results for $f= \{10^{-3}, 4\times10^{-3}, 0.04, 0.1, 0.4\}$.

In order to assess the strength of a relation between the theory error $E$ and some characteristic of the network, we calculate the coefficient of determination $R^2$ using a linear regression. For the data in Fig.~\ref{fig6}(a), we calculate $R^2 \approx 0.087$ (using the points only and ignoring the connecting curves which help identify families of points). This relatively small value indicates that $C$ is not a good predictor of the theory error across the set of networks that we tested (see Table~\ref{table1}). After examining a wide range of possibilities (see the scatter plots in Appendix~\ref{appB}), we found that the network measure that best correlates with the error $E$ (on logarithmic scales) is $(\ell - \ell_{1})/z$ (which gives $R^2 \approx 0.94$), where $z$ is the mean degree and $\ell_{1}$ is the mean intervertex distance in the version of the network that has been fully rewired while preserving the joint degree distribution $P(k,k')$ [see Fig.~\ref{fig6}(b)]. Recall that one can think of such fully $P$-rewired versions of a network as random networks with the same degree correlation $P(k,k')$ and size as the original network.

\begin{figure*}[htb]
\centering
\includegraphics[width=1.95\columnwidth]{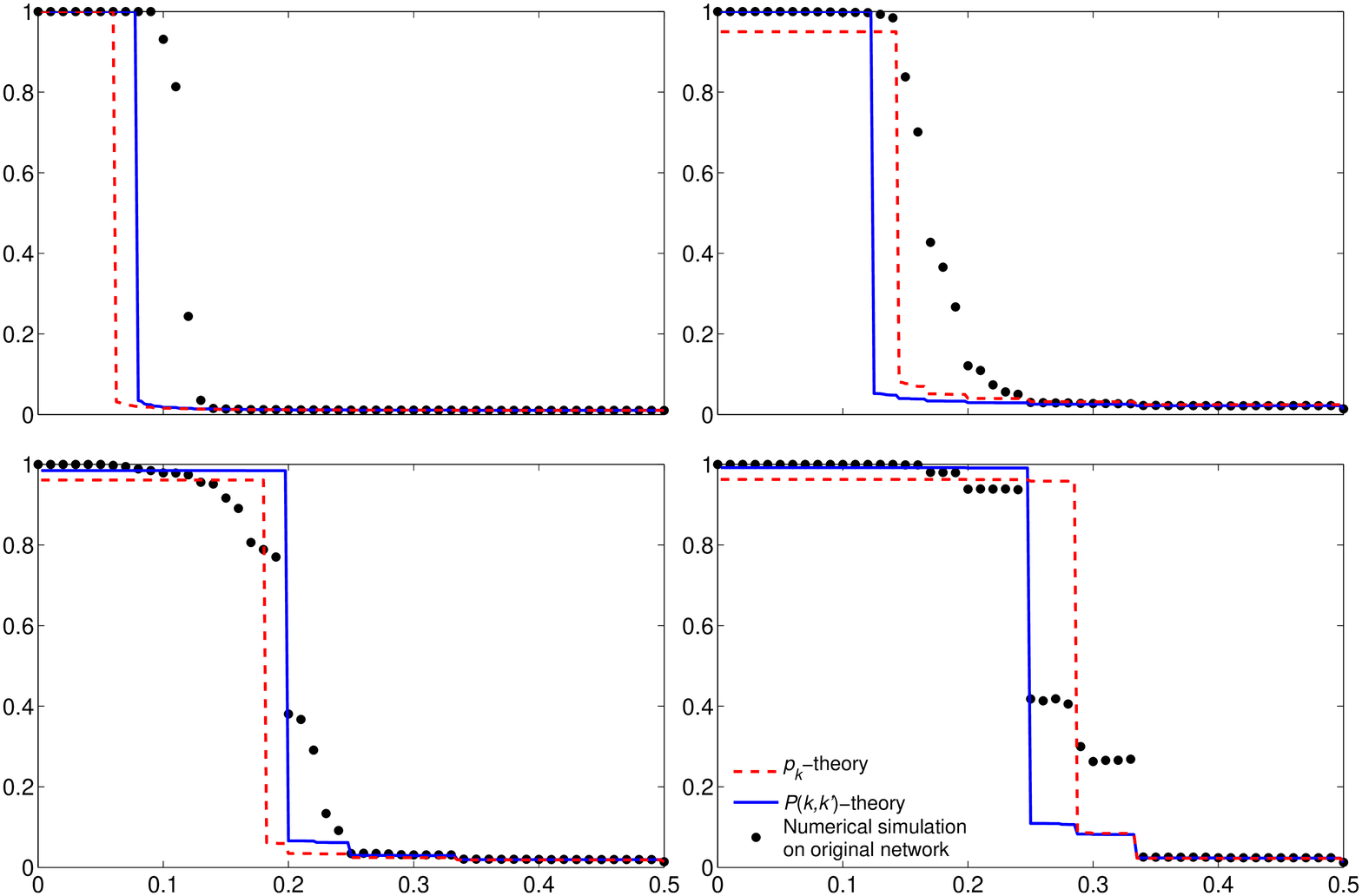}
\put(-390,280){\bf (a) Facebook Oklahoma} \put(-90,280){\bf
(b) AS Internet} \put(-360,120){\bf (c) PGP Network} \put(-90,120){\bf (d)
Power Grid} \put(-240,-15){\bf \begin{LARGE}$\mu$\end{LARGE}}
\put(-490,150){\bf \begin{LARGE}$\rho$\end{LARGE}} 
\caption{(Color online) Watts' threshold model, with threshold mean $\mu$ and variance $\sig^2 = 0$ (i.e., with uniform thresholds) for the networks from Fig.~\ref{fig1}. We use a seed fraction of $\rho_0=10^{-2}$.} 
\label{fig7}
\end{figure*}

We can summarize our observations as follows. Given a network, we compare its mean intervertex distance $\ell$ with the value $\ell_{1}$ in a random network of equal size and degree correlation $P(k,k')$. If the difference $\ell-\ell_{1}$ is sufficiently small---e.g., if it is less than $z/10$, as was the case in Fig.~\ref{fig1}(a,b)---then the $P(k,k')$-theory can be expected to accurately give the GCC size, $k$-core sizes, and results for several dynamical processes (see Figs.~\ref{fig1}--\ref{fig4}). For example, the AS Internet graph has $(\ell-\ell_1)/z \approx 3.3\times 10^{-2}$ and all 100 Facebook networks have values much smaller than this. However, the theory is not accurate for larger values of $\ell-\ell_{1}$. (For example, the PGP and Power Grid networks have $(\ell-\ell_1)/z$ values of approximately $0.45$ and $3.9$, respectively.) 

Because the tree-based theory systematically gives accurate results for dynamical processes on networks that are \emph{not} locally tree-like when the intervertex distance is small, it seems that there must be a deeper argument than is currently known for the validity of such theories. We show in Appendix~\ref{appA} that the error measure $E$ depends linearly on $\ell-\ell_1$ in a certain class of networks with zero clustering. Although this theoretical result is restricted in its applicability, it lends weight to our claim that $E$ depends primarily on $\ell-\ell_1$ rather than on the clustering $C$.

\section{Conclusions} \label{sec5}
At the beginning of this paper, we posed the following question: ``How small must small-world networks be in order for $P(k,k')$-theory to give accurate results?'' Our heuristic answer is that they must have a value for the mean intervertex distance $\ell$ that differs from the mean intervertex distance in a random network with the same $P(k,k')$ and number of nodes by no more than about 10\% of the mean degree $z$. Surprisingly, the level of clustering has much less of an impact on the accuracy of $P(k,k')$-theory, which is why we found excellent matches between theory and numerical simulations even in highly clustered graphs such as Facebook social networks and the AS Internet network.

Although our presentation used bond percolation as our primary example, we demonstrated in Figs.~\ref{fig1}--\ref{fig4} that if $P(k,k')$-theory is accurate for percolation, then it also works well for other processes. However, an absolute measure of accuracy must, of course, depend on the process under scrutiny. For example, Fig.~\ref{fig7} shows a comparison between theory and simulation results for Watts' threshold model in which $\sig=0$, which implies that all nodes have identical thresholds equal to $\mu$ (in contrast to Fig.~\ref{fig3}). This example now exhibits different results for theory and numerics even in the Facebook networks. This suggests that the $\sig=0$ case of Watts' model is particularly sensitive to deviations of the network from randomness and suggests that this case provides a suitable testing ground for new analytically solvable models of networks that include clustering~\cite{Newman09, Gleeson09a}. 

In summary, we have shown that for a variety of processes---including bond percolation and $k$-core size calculations---tree-based analytical theory yields highly accurate results for networks in which $\ell \approx \ell_{1}$ even in the presence of significant clustering. Such graphs, which include the AS Internet network and Facebook social networks, are definitively not locally tree-like, so that the theory is working very well even in situations where the theory's fundamental hypothesis is known to fail utterly. The fact that analytical results for several dynamical processes can be expected to apply on ``sufficiently small'' small-world networks increases the value of existing theoretical work and highlights the types of process for which improved analytical modelling of clustering effects should most profitably be targeted. We hope that the results of the present paper will motivate further research on the underlying causes of this ``unreasonable'' effectiveness of tree-based theory for clustered networks.

\section*{Acknowledgements}
SM, AH, and JPG acknowledge funding provided by Science Foundation Ireland under programmes 06/IN.1/I366 and MACSI 06/MI/005. MAP acknowledges a research award (\#220020177) from the James S. McDonnell Foundation. PJM was funded by the NSF (DMS-0645369). We thank Adam D'Angelo and Facebook for providing the Facebook data used in this study. We also thank Alex Arenas, Mark Newman, CAIDA, and Cx-Nets collaboratory for making publicly available other data sets used in this paper.

\appendix
\section{Scaling of Prediction Error with Mean Intervertex Distance} \label{appA}
We consider the class of networks for which one can define a \emph{branching matrix}~\cite{Goltsev08}. A branching matrix describes the connection probabilities in tree-like networks with non-trivial structure, e.g., modular networks~\cite{Dorogovtsev08b}. In this appendix, we derive how the error measure $E$ defined in Eq.~\eqref{e2} depends on $\ell-\ell_1$ for a network with a branching matrix when the network is close to fully $P$-rewired (i.e., when it is close to a random network with the same degree correlation). We give the final formula in Eq.~(\ref{e8}) below. Because clustering is negligible in these infinite networks, $E$ cannot depend on the clustering coefficient $C$. In Fig.~\ref{fig6}, we illustrate both of these characteristics for real-world networks.

The branching matrix characterizes the average intervertex distance $\ell$ in a network, and it also determines the bond percolation behavior. The largest eigenvalue of the branching matrix, which we denote by $\lam$, determines the percolation threshold
\begin{equation}
p_{\rm th} = \frac{1}{\lam}\,. \label{e3}
\end{equation}
Additionally, an estimate of the mean intervertex distance can be written in terms of $\lam$ as~\cite{Goltsev08}
\begin{equation}
\ell \approx \frac{\ln N}{\ln \lam}\,, \label{e4}
\end{equation}
where we recall that $N$ denotes the number of nodes in the network.

We suppose now that the network is almost fully $P$-rewired, and we consider how values of $\lam$ that differ from the fully $P$-rewired value (which we denote by $\lam_1$) affect the values of $\ell$ and $p_{\rm th}$. Note that it is easy to calculate $\lam_1$, as the branching matrix of a fully $P$-rewired network is given in terms of the degree correlation matrix $P(k,k^\prime)$ by~\cite{Goltsev08}
\begin{equation}
B_1(k,k^\prime) \equiv (k^\prime-1) \frac{P(k,k^\prime)}{\sum_{j} P(k,j)} 
=(k^\prime-1) \frac{P(k,k^\prime)}{k p_k/z}\,, \label{e5}
\end{equation}
and $\lam_1$ is the largest eigenvalue of $B_1$. Moreover, for uncorrelated networks produced using the configuration model, $\lam_1$ is simply $\sum_k k(k-1)p_k/z$. This implies in particular that $\lam_1 = z-1$ for graphs in which all nodes have the same degree (such as $P$-rewired Watts-Strogatz networks and the special cases of $\g$-theory networks used in Sec.~\ref{sec3}).

Considering only small deviations from fully $P$-rewired values, we write $\lam = \lam_1 + \D \lam$, and $\ell = \ell_1 +\D\ell$. Expanding to linear terms, we find from~\eqref{e4} that the excess length is
\begin{equation}
\D\ell= -\frac{\D\lam\, \ln N}{\lam_1(\ln\lam_1)^2}\,. 
\label{e6}
\end{equation}
Similarly, we find from (\ref{e3}) that the change in percolation threshold is
\begin{equation}
\D p_{\rm th} = -\frac{\D \lam}{\lam_1^2}\,.
\label{e7}
\end{equation}
If we now make the further assumption that $\D p_{\rm th}$ is approximately equal to the error $E$ for the bond percolation process [this approximation is exact if the effect of the perturbation is to shift the entire bond percolation curve $S(p)$ to $S(p+\D p_{\rm th})$], we obtain the relation
\begin{equation}
E \approx \frac{(\ln \lam_1)^2}{\lam_1 \ln N} (\ell-\ell_1)\,.
\label{e8}
\end{equation}

\begin{figure}[ht]
\centering
\includegraphics[width=0.95\columnwidth]{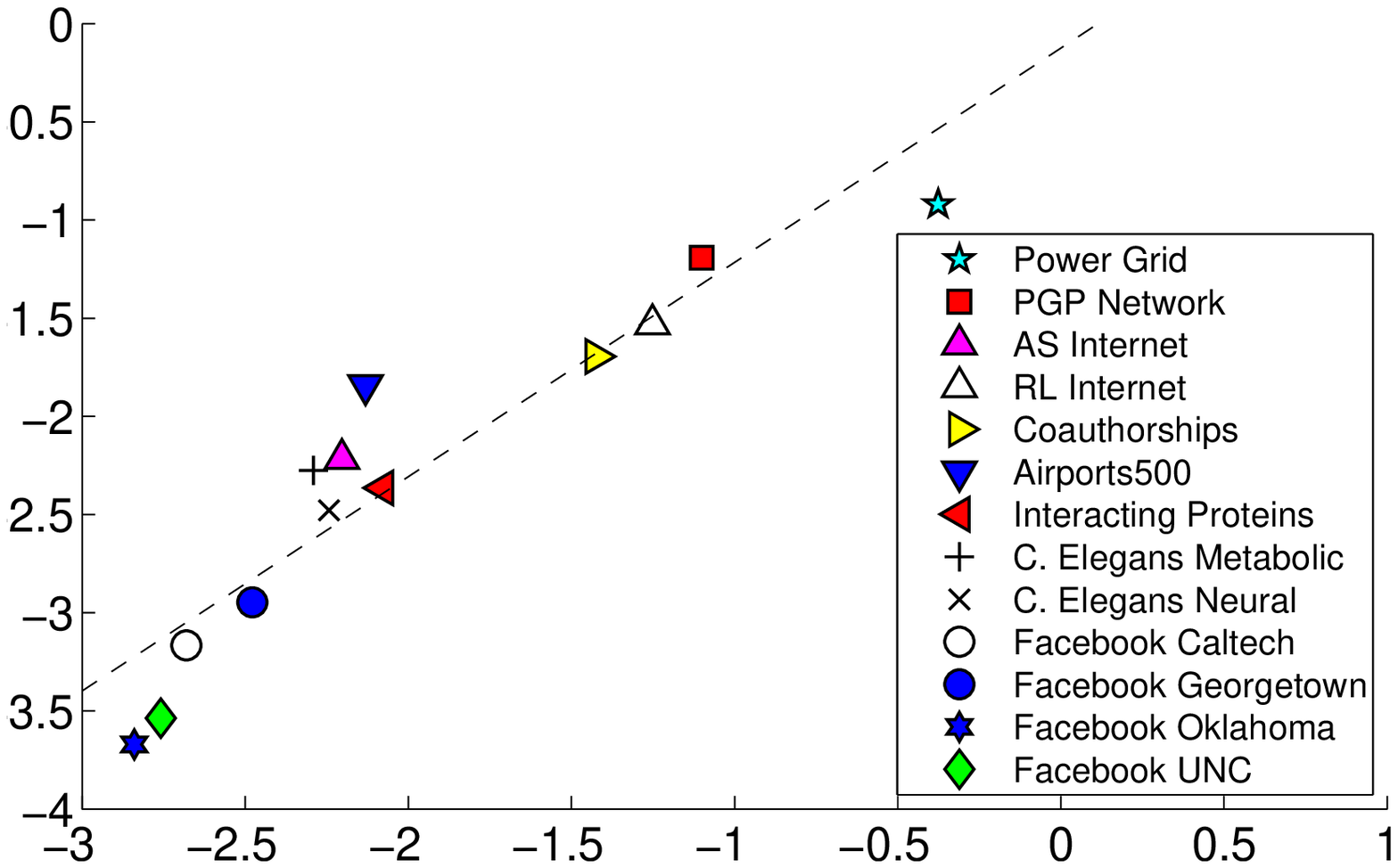}
\put(-160,-15){\bf \begin{large}$\log_{10}[\frac{(\ln \lam_1)^2}{\lam_1 \ln N} (\ell-\ell_1)]$\end{large}}
\put(-240,65){\begin{rotate}{90} \bf\begin{large}$\log_{10} E$\end{large} \end{rotate}}
\caption{(Color online) Log-log scatter plot of actual (numerical) values of $E$ for real-world networks versus the values predicted by Eq.~\eqref{e8}, for which we numerically calculate $\ell$ and $\ell_1$. We find that $R^2\approx 0.87$; the slope of the fitted line is 1.09.
} 
\label{fig8}
\end{figure}

Although the scope of our analysis is obviously limited by our assumptions, Eq.~\eqref{e8} nevertheless supports our main claim that $E$ depends primarily on the excess length $\ell-\ell_1$. Note $C=0$ for branching-matrix networks, so $E$ is (trivially) independent of $C$; compare this to the results for the real-world networks that are shown in Fig.~\ref{fig6}(a). Moreover, the scatter plot of $\log_{10}E$ versus $\log_{10}[(\ln \lam_1)^2 (\ell-\ell_1)/(\lam_1 \ln N)]$ in Fig.~\ref{fig8} indicates that Eq.~\eqref{e8} gives a good fit ($R^2 \approx 0.87$) even for real-world networks.

\section{Scatter Plots} \label{appB}
\setlength{\tabcolsep}{0.25cm}

\begin{table*}[floatfix]
\begin{center}
\begin{tabular}{l|l|r|r|r|r|l|c|c|r|l}
\hline
&Network    	&   $N$ & $z$ 	&$\ell$	&$\ell_1$& $\ell_1^{B}$	& $C$	& $\WT C$ & $r$ & Ref(s).\\
\hline
\begin{rotate}{90}
\hspace{-3.5cm}Real world
\end{rotate}
&Power Grid     	& 4941  & 2.67	& 18.99 & 8.61  &  7.85 & 0.08  & 0.10   & 0.0035 	& \cite{Watts98, power_url}\\
&PGP Network    	& 10680 & 4.55	& 7.49  & 5.40  &  2.66 & 0.27  & 0.38 	 & 0.23		& \cite{Guardiola02, Boguna04, PGP_url}\\
&AS Internet    	& 28311 & 4.00	& 3.88  & 3.67  &  2.56 & 0.21  & 0.0071 & -0.20	& \cite{asrel_url}\\
&RL Internet    	& 190914& 6.34  & 6.98  & 5.25  &  3.17 & 0.16  & 0.061	 & 0.025	& \cite{ITDK_url}\\
&Coauthorships  	& 39577 & 8.88	& 5.50  & 4.45  &  2.93 & 0.65  & 0.25	 & 0.19		& \cite{Newman01b, cond_mat_url}\\
&Airports500     	& 500   & 11.92 & 2.99	& 2.76	&  1.62 & 0.62  & 0.35	 & -0.278	& \cite{Colizza07, air500_url}\\
&Interacting Proteins   & 4713  & 6.30	& 4.22	& 4.05	&  2.96 & 0.09  & 0.062	 & -0.136	& \cite{Colizza05,Colizza06,DIP_url}\\
&C. Elegans Metabolic   & 453	& 8.94	& 2.66	& 2.55	&  1.93 & 0.65  & 0.12	 & -0.226	& \cite{Duch05,cel_met_url}\\
&C. Elegans Neural   	& 297	& 14.46	& 2.46  & 2.33	&  1.84 & 0.29  & 0.18	 & -0.163	& \cite{Watts98,cel_neur_url}\\
&Facebook Caltech     	& 762   & 43.70	& 2.34  & 2.26  &  1.55 & 0.41  & 0.29	 & -0.066	& \cite{Traud08}\\
&Facebook Georgetown  	& 9388  & 90.67	& 2.76  & 2.55  &  1.79 & 0.22  & 0.15	 & 0.075	& \cite{Traud08}\\
&Facebook Oklahoma    	& 17420 &102.47	& 2.77  & 2.66  &  1.79 & 0.23	& 0.16	 & 0.074	& \cite{Traud08}\\
&Facebook UNC         	& 18158 & 84.46	& 2.80  & 2.68  &  1.87 & 0.20  & 0.12	 & 7$\times10^{-5}$	& \cite{Traud08}\\
\hline
\begin{rotate}{90}
\hbox{\hspace{-1.3cm}Synthetic}
\end{rotate}
&$\g$-theory [$\g(3,3)=1]$& 1002  & 3	& 13.15 & 8.06  &   9.97  & 1/3 & 1/3    & N/A & \cite{Gleeson09a}\\
&$\g$-theory [$\g(3,3)=1]$&10002  & 3	& 19.81 & 11.37 &  13.29  & 1/3 & 1/3    & N/A & \cite{Gleeson09a}\\
&Watts-Strogatz (WS)      & 1000  & 10	& 50.45	& 3.29  &   3.14  & 2/3 & 2/3    & N/A & \cite{Watts98}\\
&Watts-Strogatz (WS)      &10000  & 10	& 500.45& 4.34  &   4.19  & 2/3 & 2/3    & N/A & \cite{Watts98}\\
\end{tabular}
\end{center}
\caption{Basic summary statistics for the networks that we used in this paper. We have treated all real-world data sets as undirected, unweighted networks and have computed the following properties: total number of nodes $N$; mean degree $z$; mean intervertex distance $\ell$ in original network; mean intervertex distance $\ell_1$ in the corresponding fully $P$-rewired version of the network (i.e., in a random network with the original degree correlation); the mean intervertex distance $\ell_1^{B}$ predicted by Eq.~\eqref{e4} using the branching matrix corresponding to a random network with the original degree correlation; clustering coefficients $C$ and $\WT C$ (whose respective definitions are given by Eqs.~(3.6) and (3.4) of~\cite{Newman03a}); and the Pearson degree correlation coefficient $r$. The last column in the table gives the citation number(s) for the data in the bibliography. 
}
\label{table1}
\end{table*}

In this appendix, we show scatter plots of $\log_{10} E$ versus a variety of possible predictors. Recall that $E$, which we defined in Eq.~(\ref{e2}), gives an error measure for bond percolation. We test for the dependence of $E$ on various combinations of the mean degree $z$, mean intervertex distance $\ell$, and clustering coefficients ~\footnote{We consider both common definitions of clustering coefficient. We use $C$ to denote the coefficient defined by Eq.~(3.6) of Ref.~\cite{Newman03a} and $\WT C$ to denote that from Eq.~(3.4) of Ref.~\cite{Newman03a}.}. Recall again that $\ell_1$ denotes the value taken by $\ell$ in a fully $P$-rewired version of a network (i.e., in a random network with the same degree correlation and size).

The scatter plots show data points for real-world networks, and for synthetic Watts-Strogatz small-world networks and $\g$-theory networks, which are described in Sec.~\ref{sect3B}. The dependence of $E$ on $\ell-\ell_1$ is clearly strong (see the top row of scatter plots, which all have $R^2>0.9$), whereas the dependence on clustering is weak (see the bottom row of scatter plots, which all have $R^2<0.3$). Given the relatively small number of available data sets, we cannot definitively select the best scaling function $F(z,\ell,\ldots)$ for the relation $E \approx F(z,\ell,\ldots) (\ell-\ell_1)$, but the simple choice $F=1/z$ used in Fig.~\ref{fig6}(b) and the scaling function $F=\ln^2 \lam_1/(\lam_1 \ln N)$ indicated by Eq.~\eqref{e8} both give satisfactory fits.

\def\putpan{\put(-144,102)}
\def\putx{\put(-95,-10)}
\def\hs{\hspace{0.1cm}}
\def\vs{\vspace{0.2cm}}
\def\includegraphicsWH{\includegraphics[width=0.65\columnwidth]}
\begin{figure*}[ht]
\begin{flushright}
\includegraphicsWH{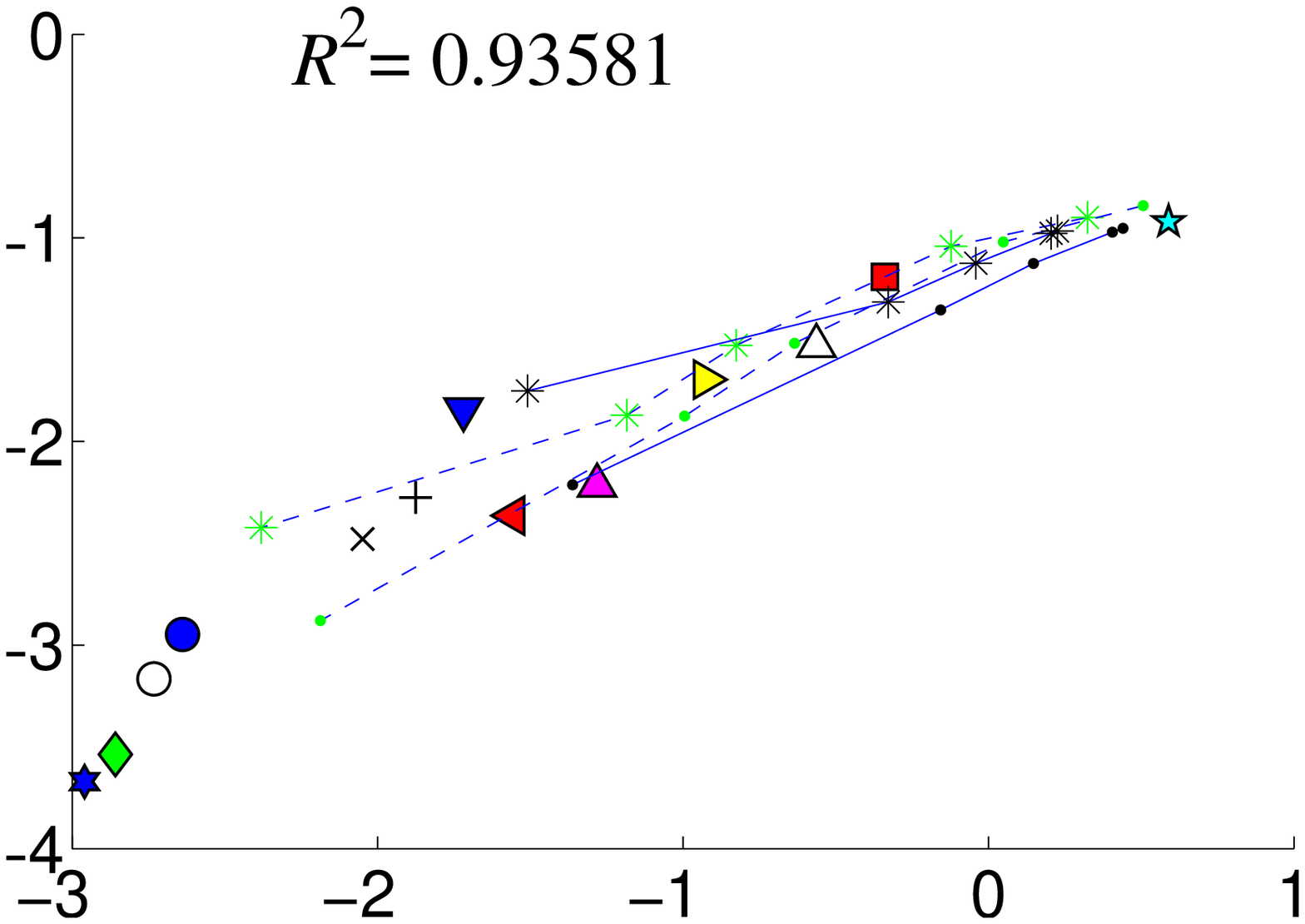} \putpan{\bf(a)} \putx{\bf$\log_{10}[(\ell-\ell_1)/z]$} \hs 
\includegraphicsWH{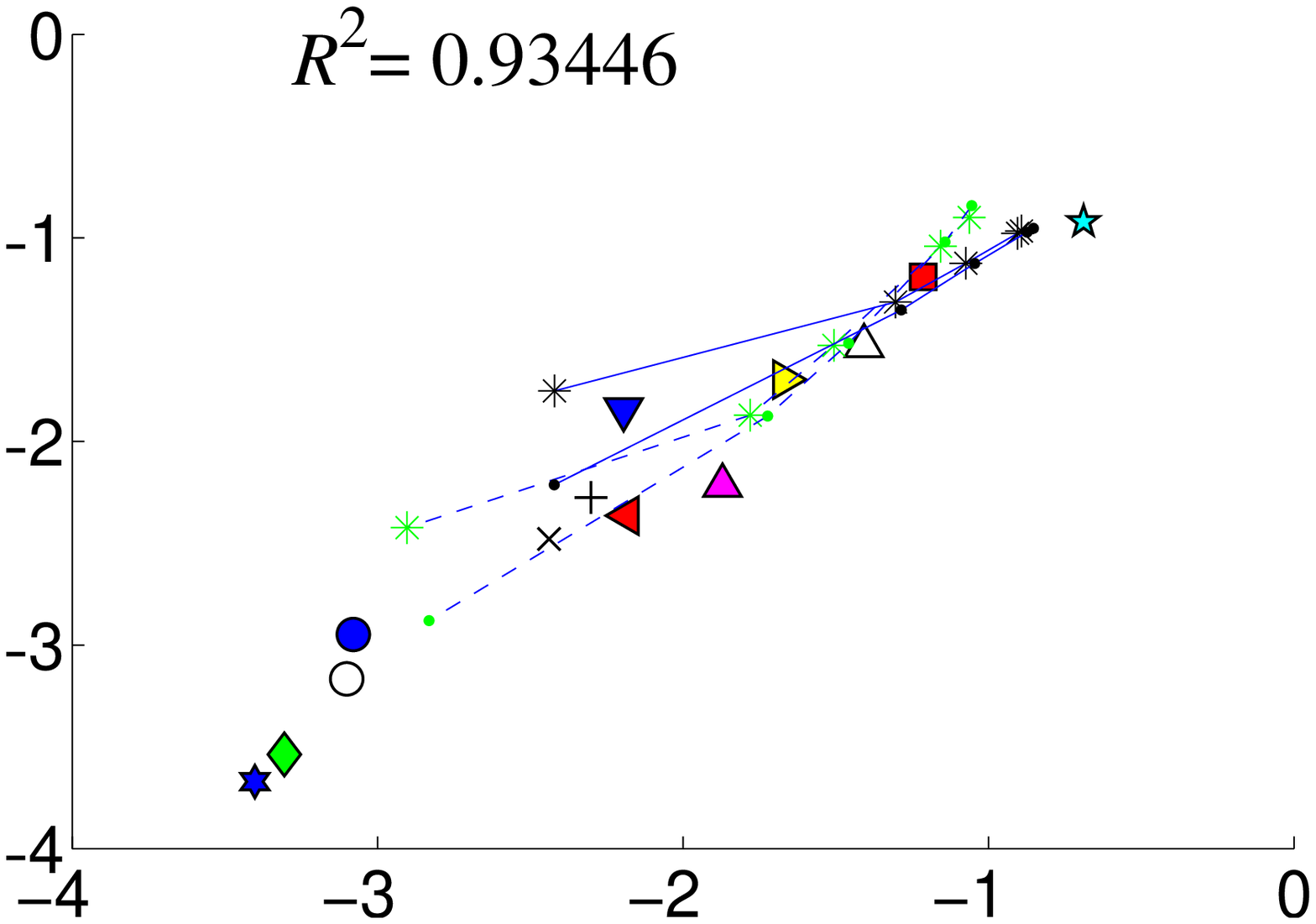} \putpan{\bf(b)} \putx{\bf$\log_{10}[(\ell-\ell_1)/(\ell z)]$} \hs 
\includegraphicsWH{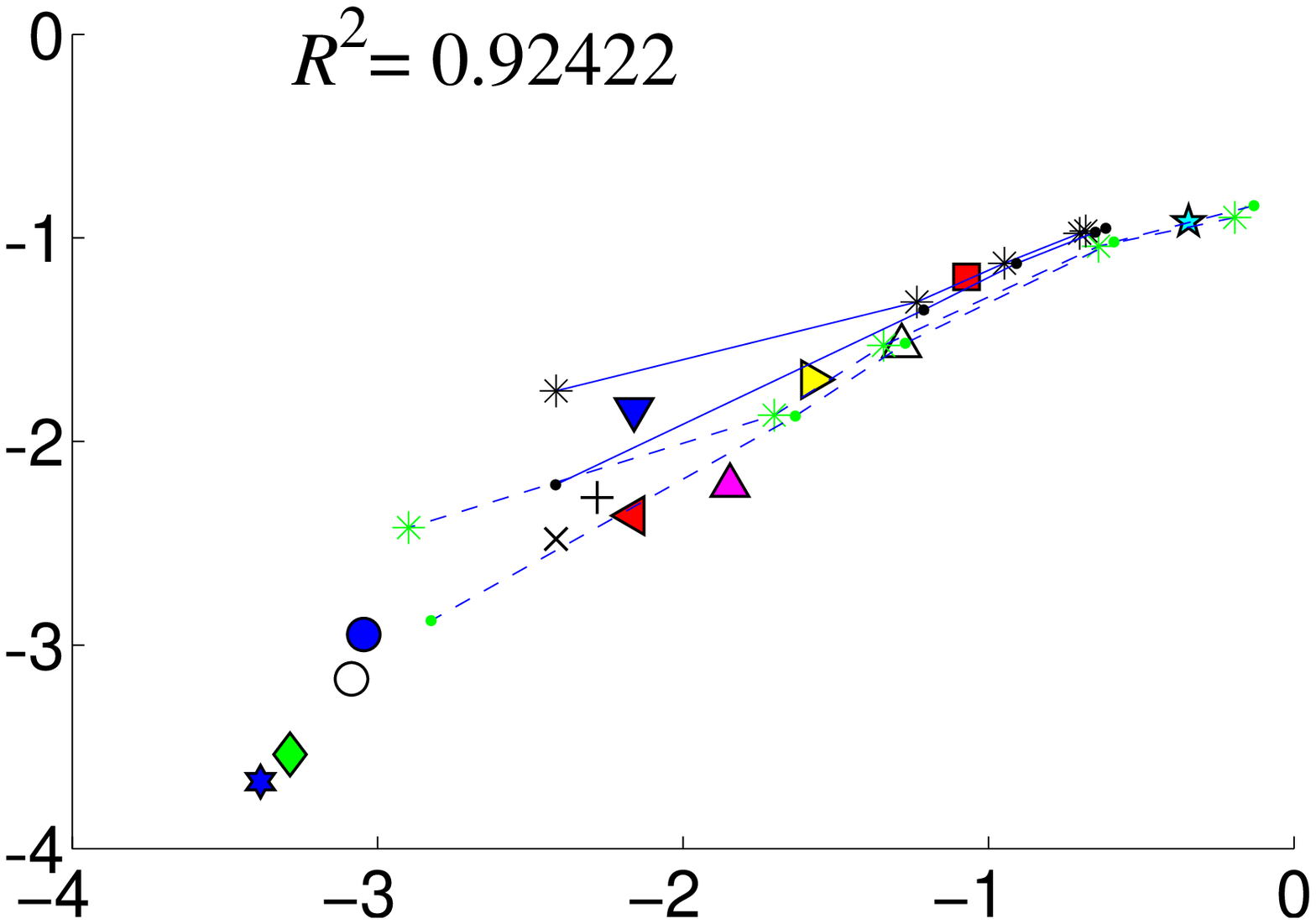} \putpan{\bf(c)} \putx{\bf$\log_{10}[(\ell-\ell_1)/(\ell_1 z)]$}
\\ \vs
\includegraphicsWH{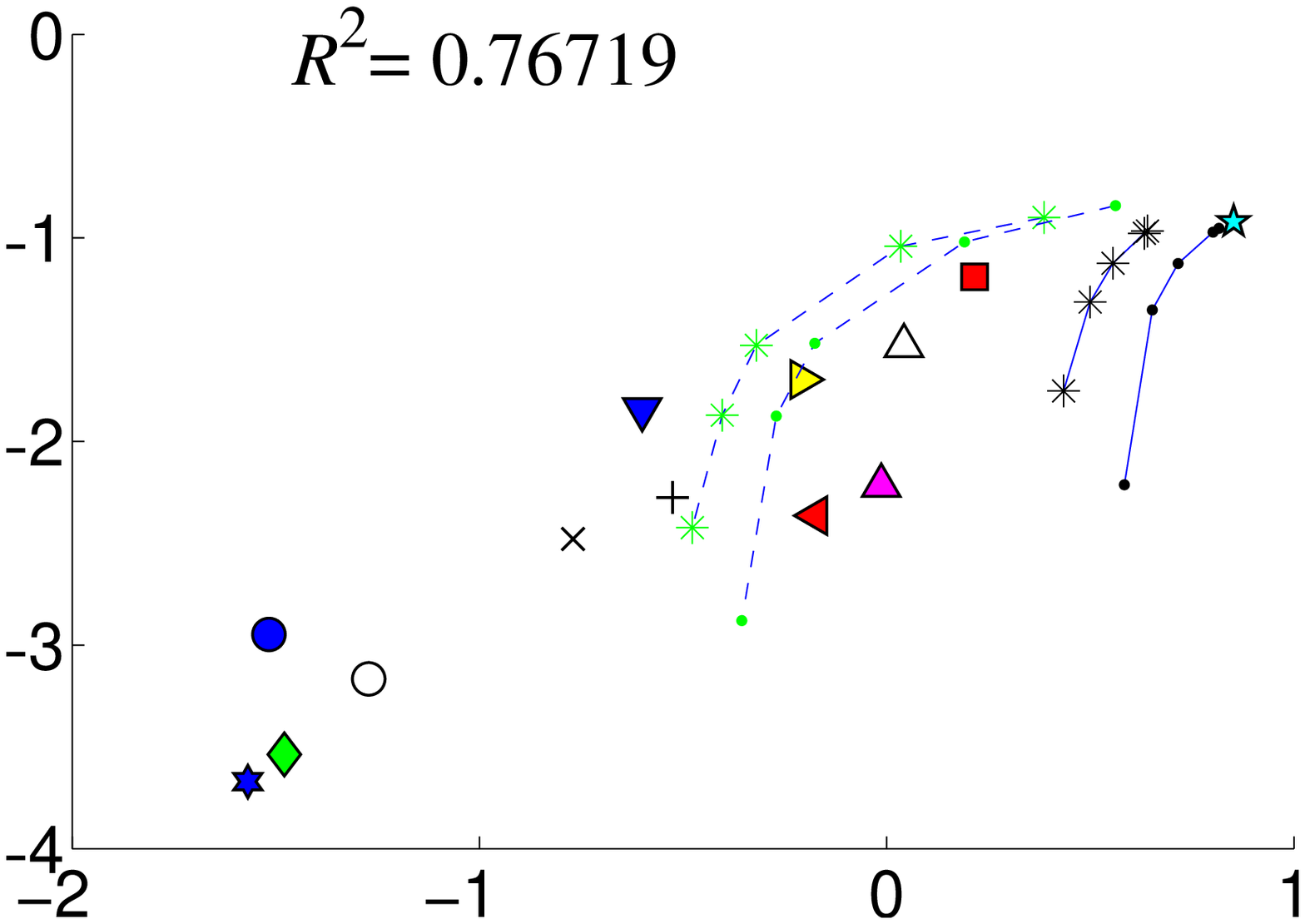} \putpan{\bf(d)} \putx{\bf$\log_{10}(\ell/z)$} \hs 
\includegraphicsWH{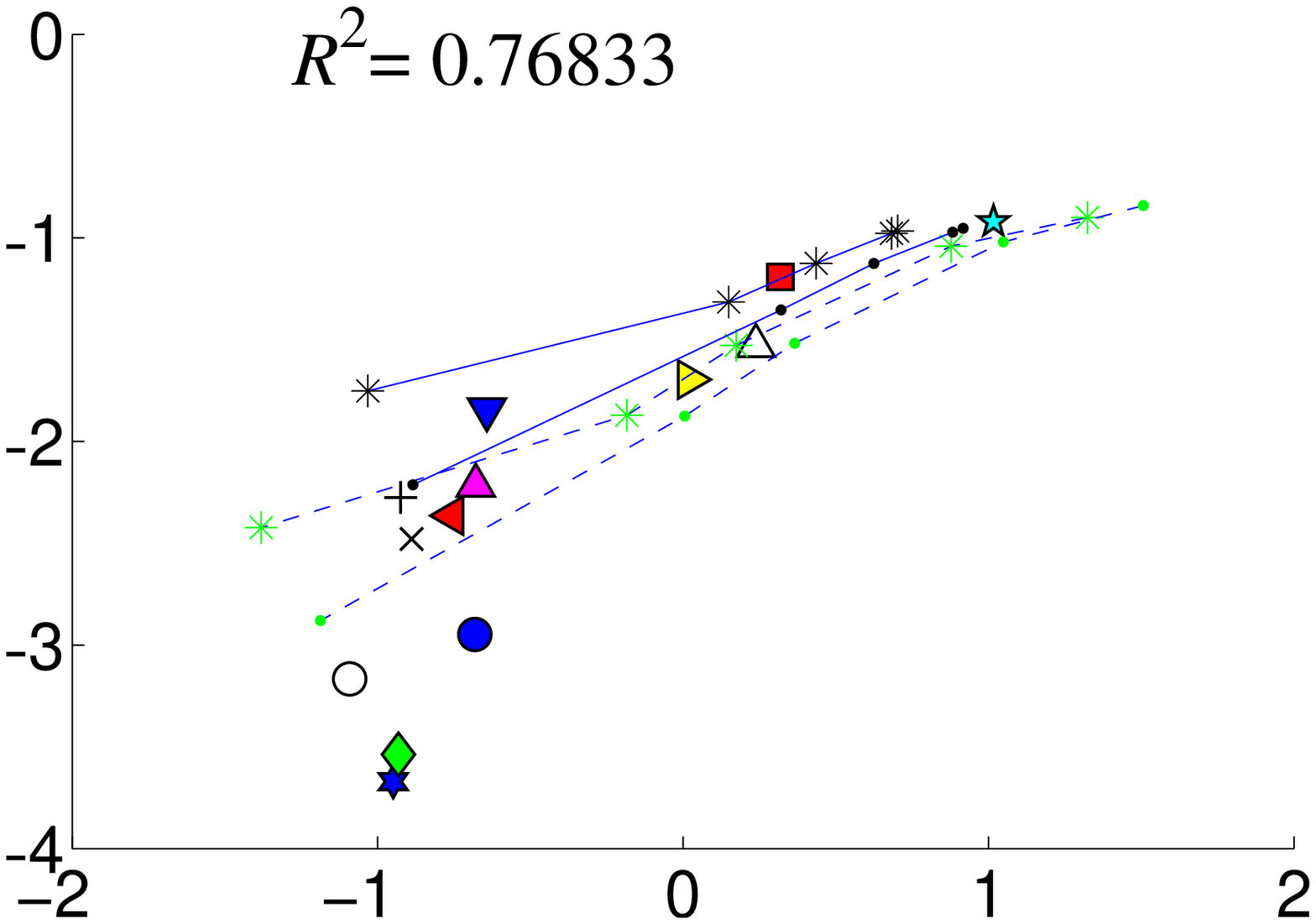} \putpan{\bf(e)} \putx{\bf$\log_{10}(\ell-\ell_1)$} \hs 
\includegraphicsWH{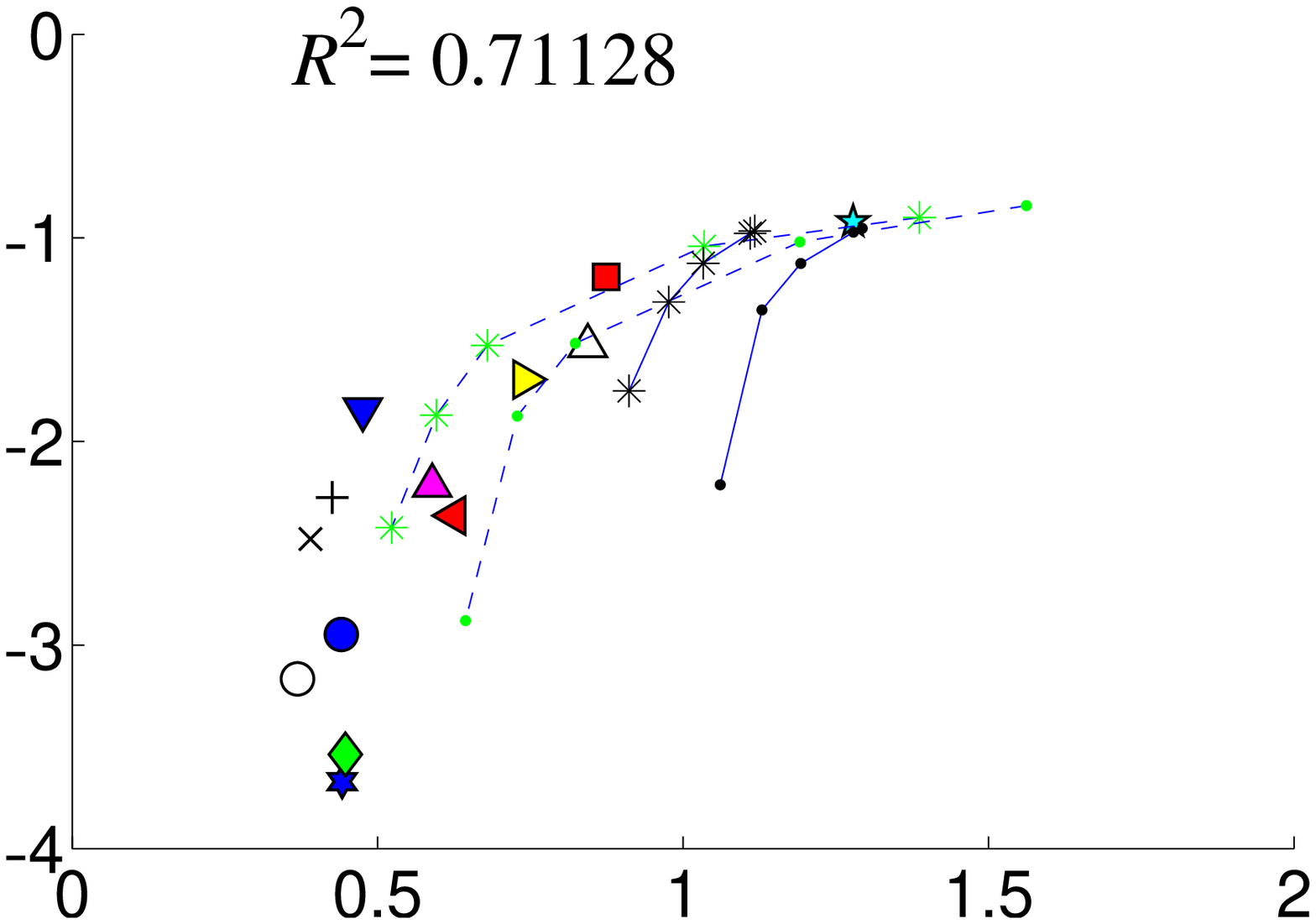} \putpan{\bf(f)} \putx{\bf$\log_{10}(\ell)$}
\\ \vs
\includegraphicsWH{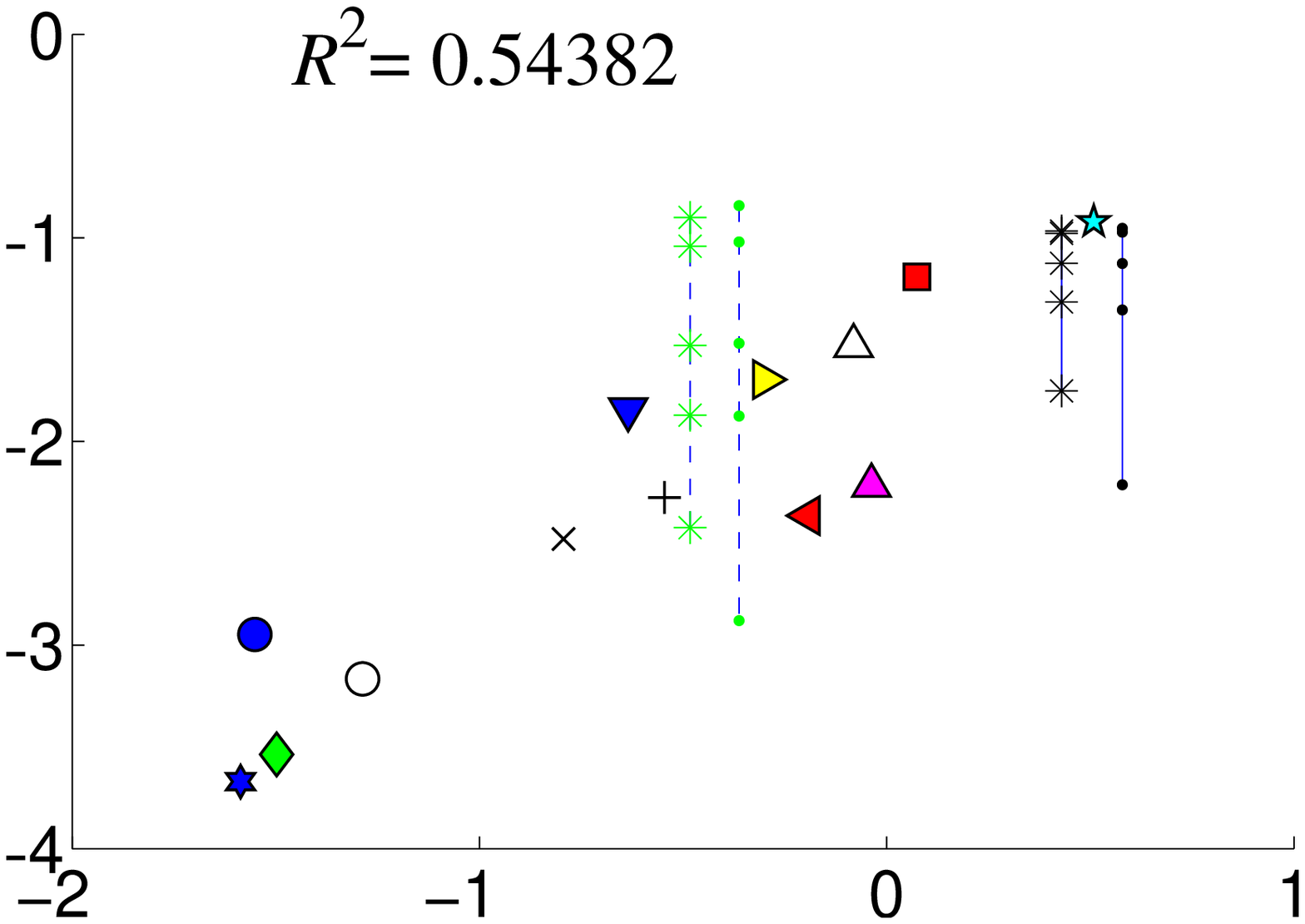} \putpan{\bf(g)} \putx{\bf$\log_{10}(\ell_1/z)$} \hs 
\includegraphicsWH{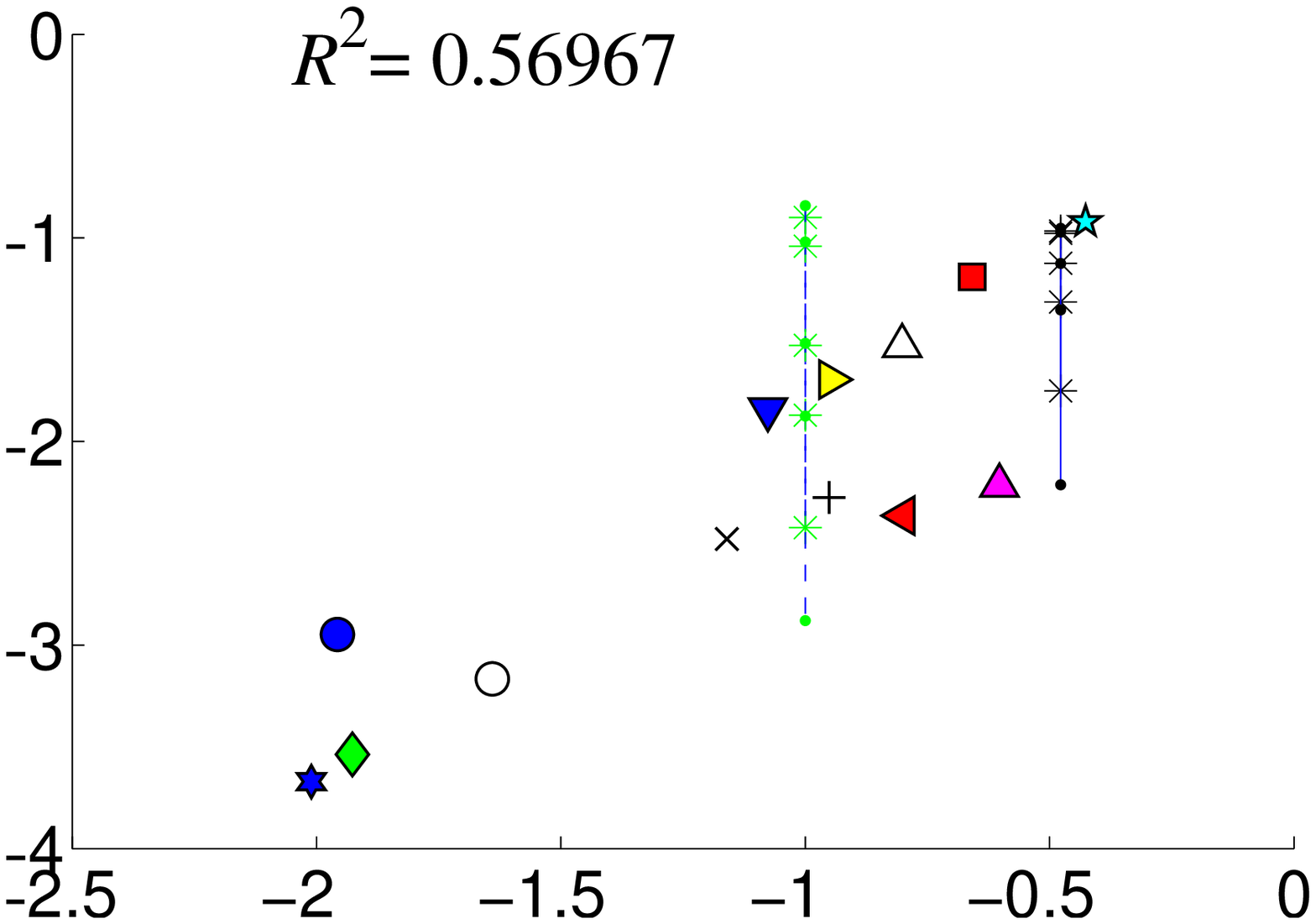} \putpan{\bf(h)} \putx{\bf$\log_{10}(1/z)$} \hs 
\includegraphicsWH{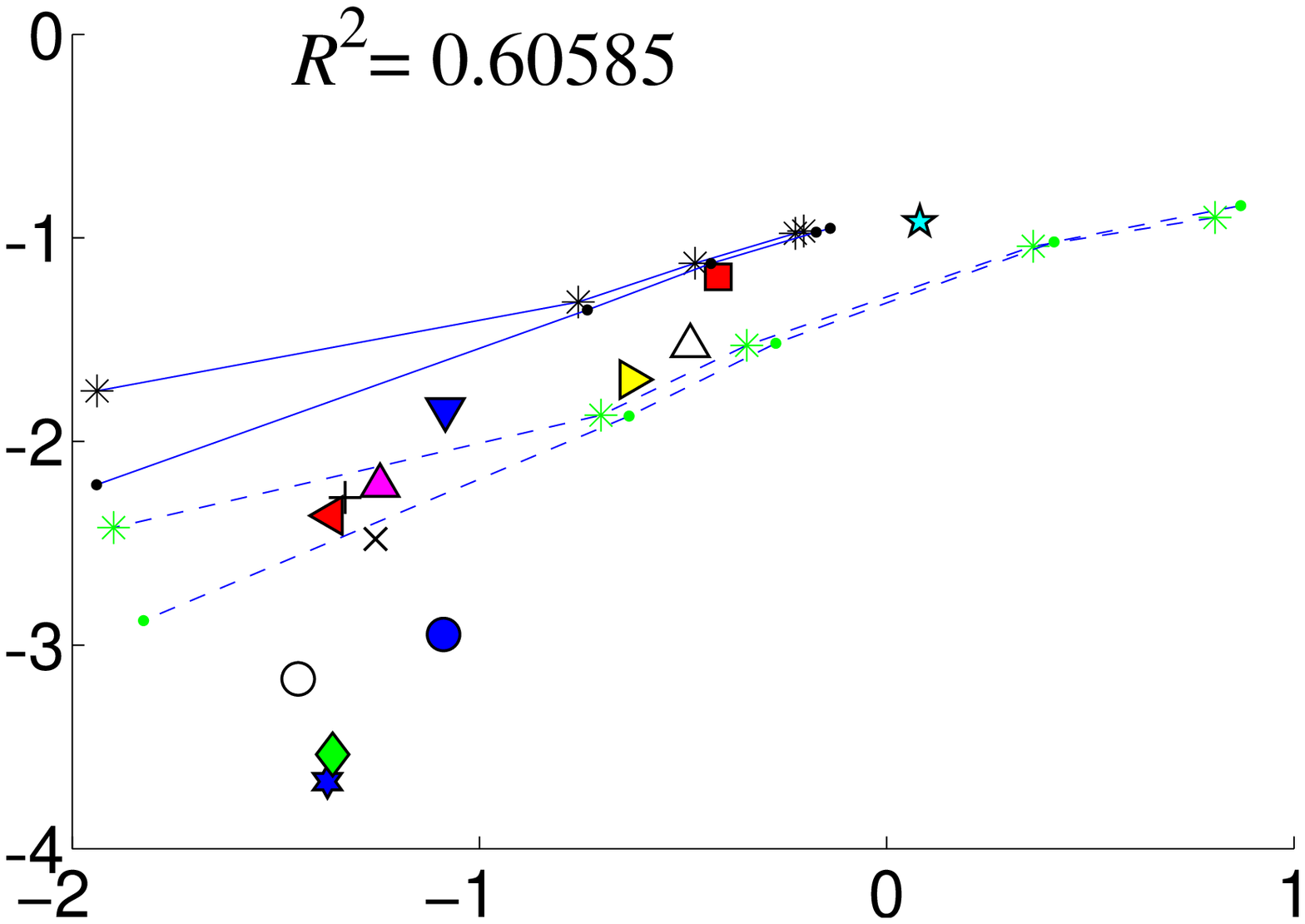} \putpan{\bf(i)} \putx{\bf$\log_{10}[(\ell-\ell_1)/\ell_1]$}
\\ \vs
\includegraphicsWH{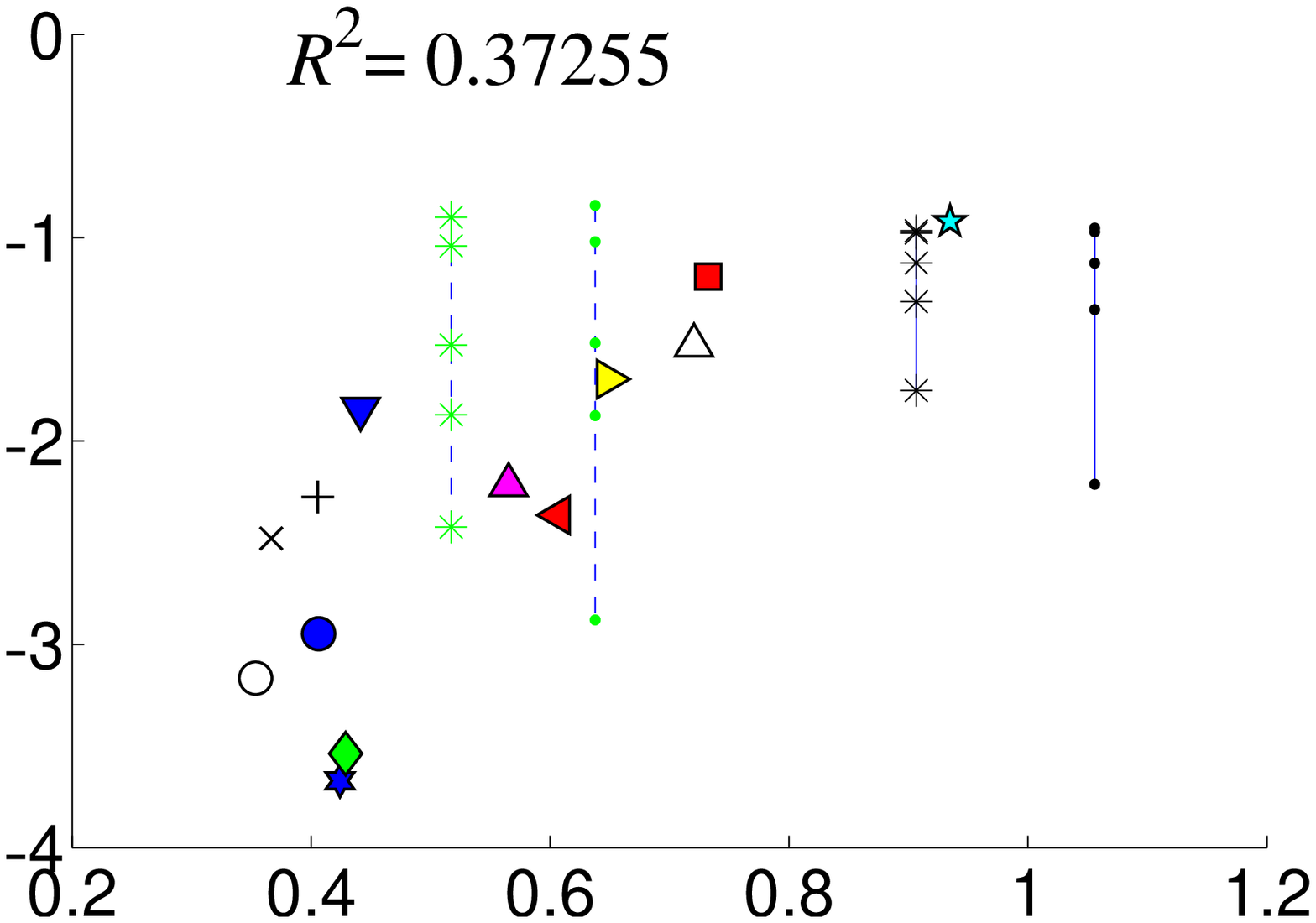} \putpan{\bf(j)} \putx{\bf$\log_{10}(\ell_1)$} \hs 
\includegraphicsWH{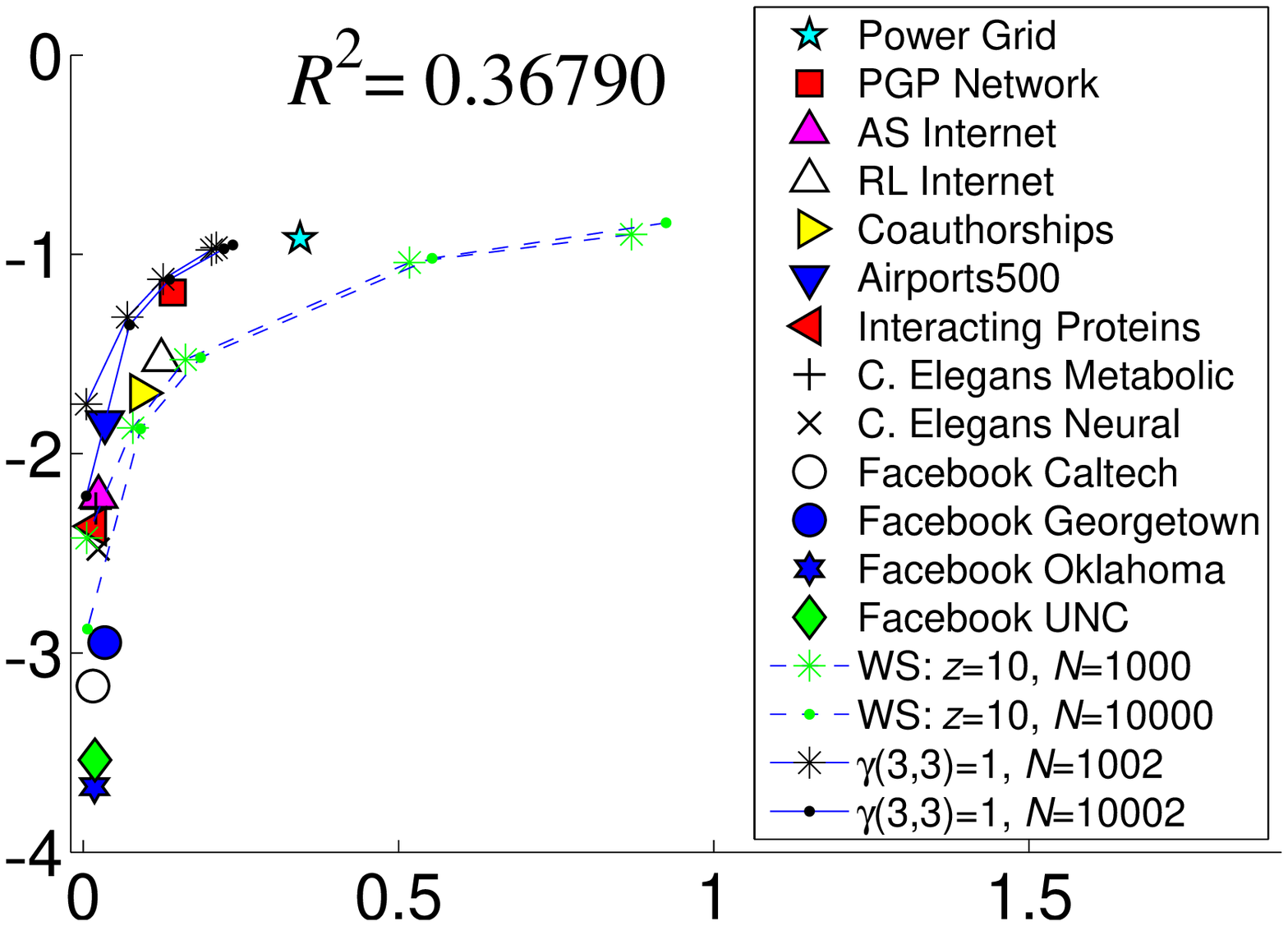} \putpan{\bf(k)} \putx{\bf$\log_{10}(\ell/\ell_1)$} \hs 
\includegraphicsWH{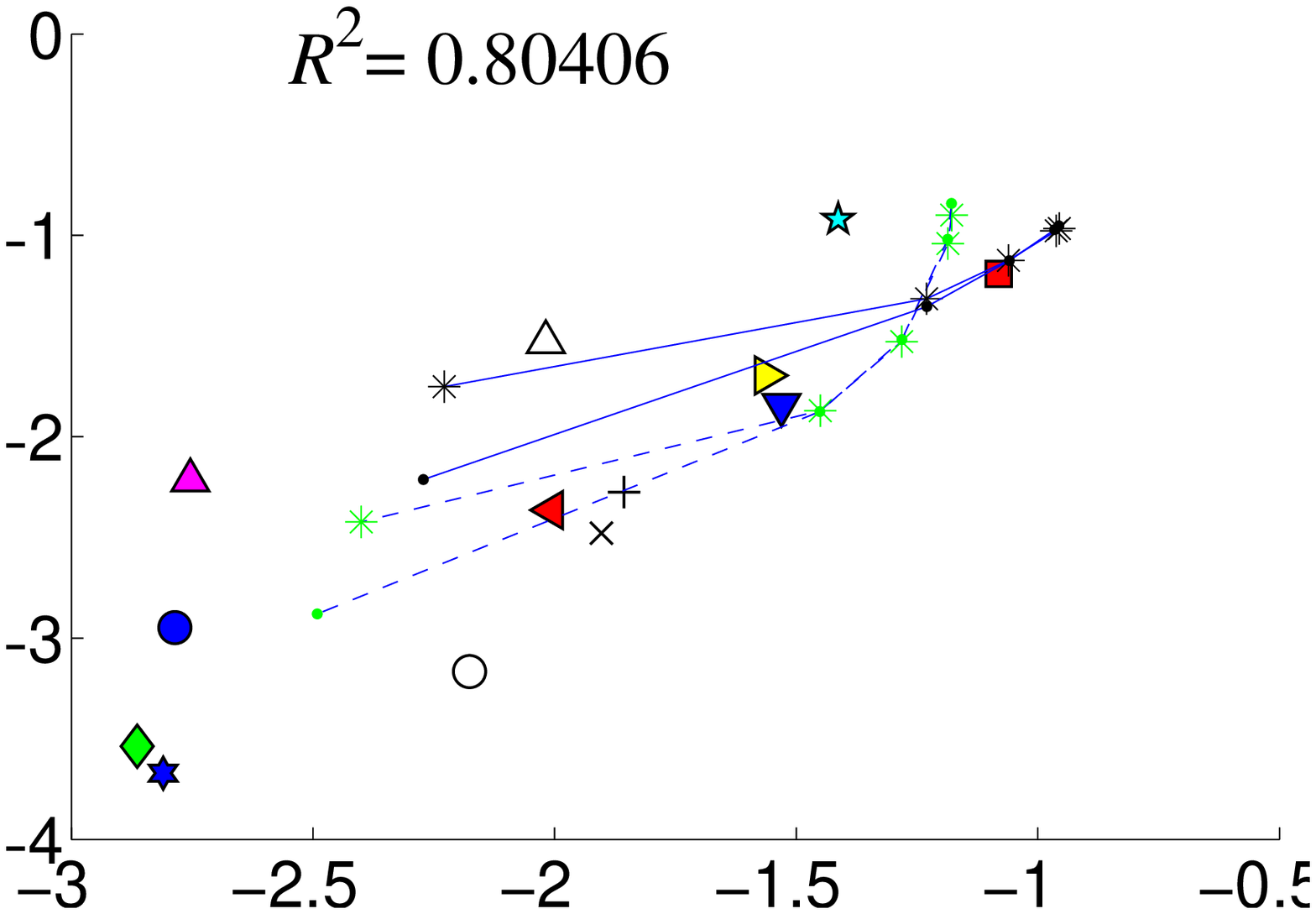} \putpan{\bf(l)} \putx{\bf$\log_{10}(\WT C / z)$}
\\ \vs
\includegraphicsWH{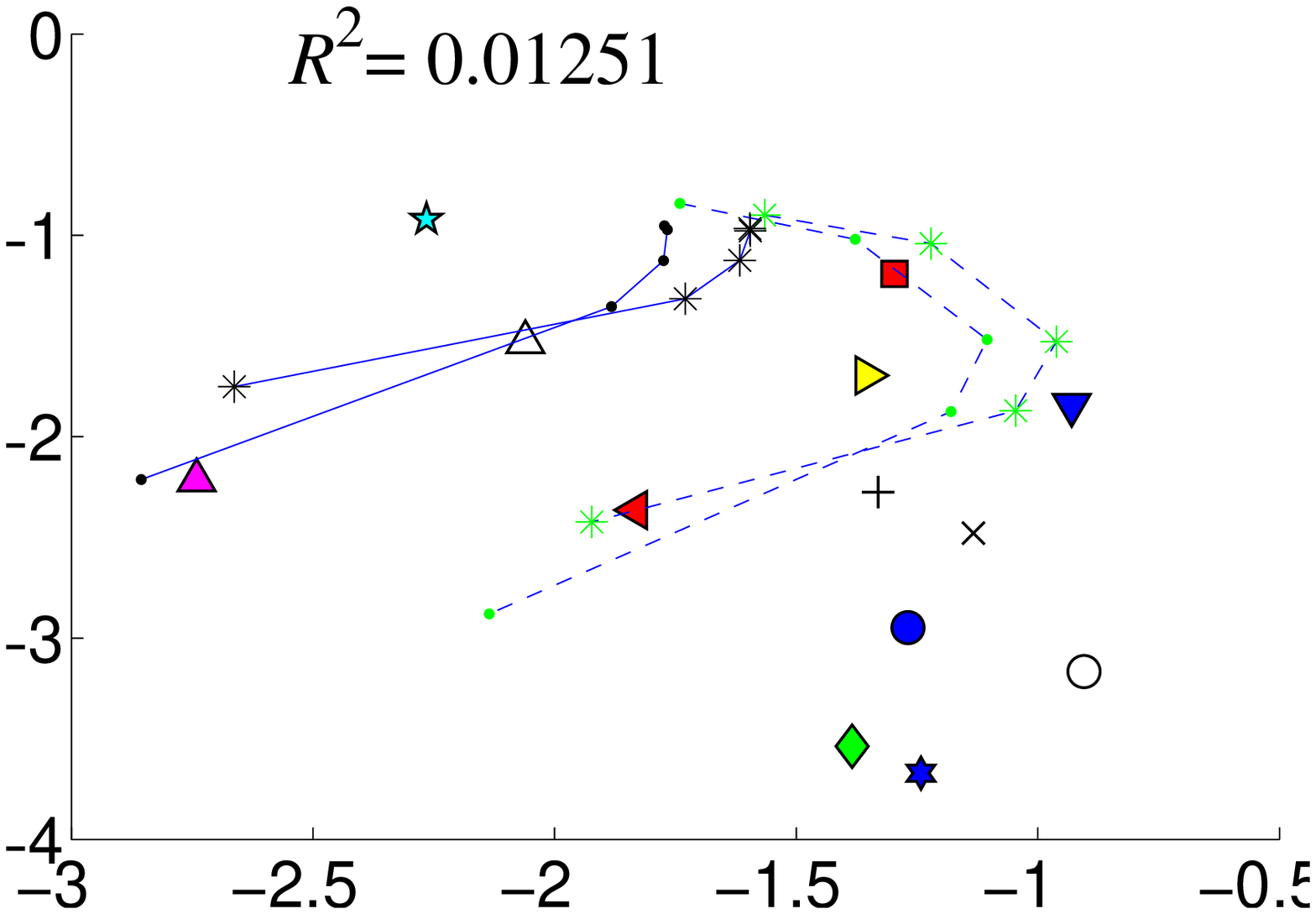} \putpan{\bf(m)} \putx{\bf$\log_{10}(\WT C / \ell)$} \hs 
\includegraphicsWH{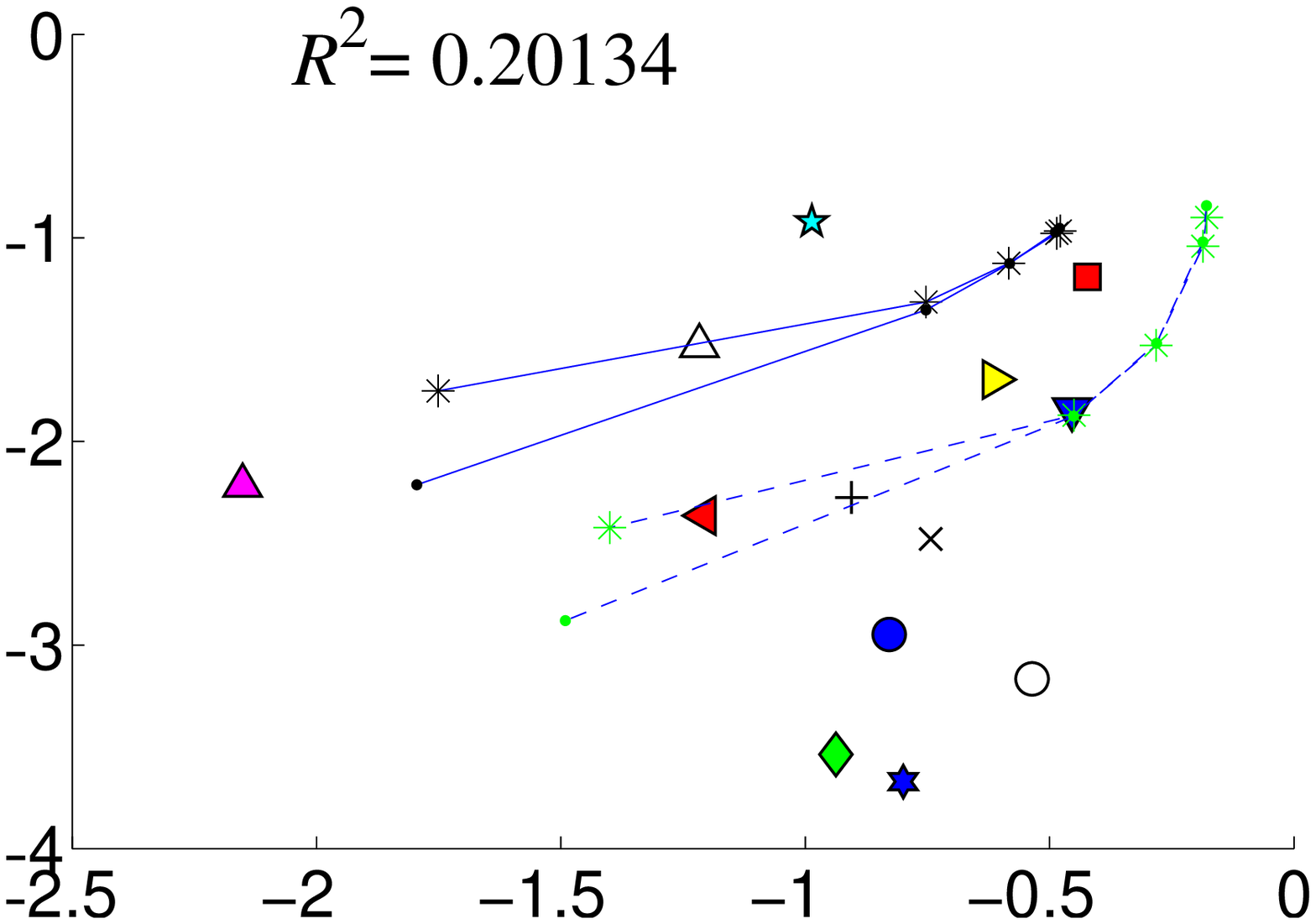} \putpan{\bf(n)} \putx{\bf$\log_{10}(\WT C)$} \hs 
\includegraphicsWH{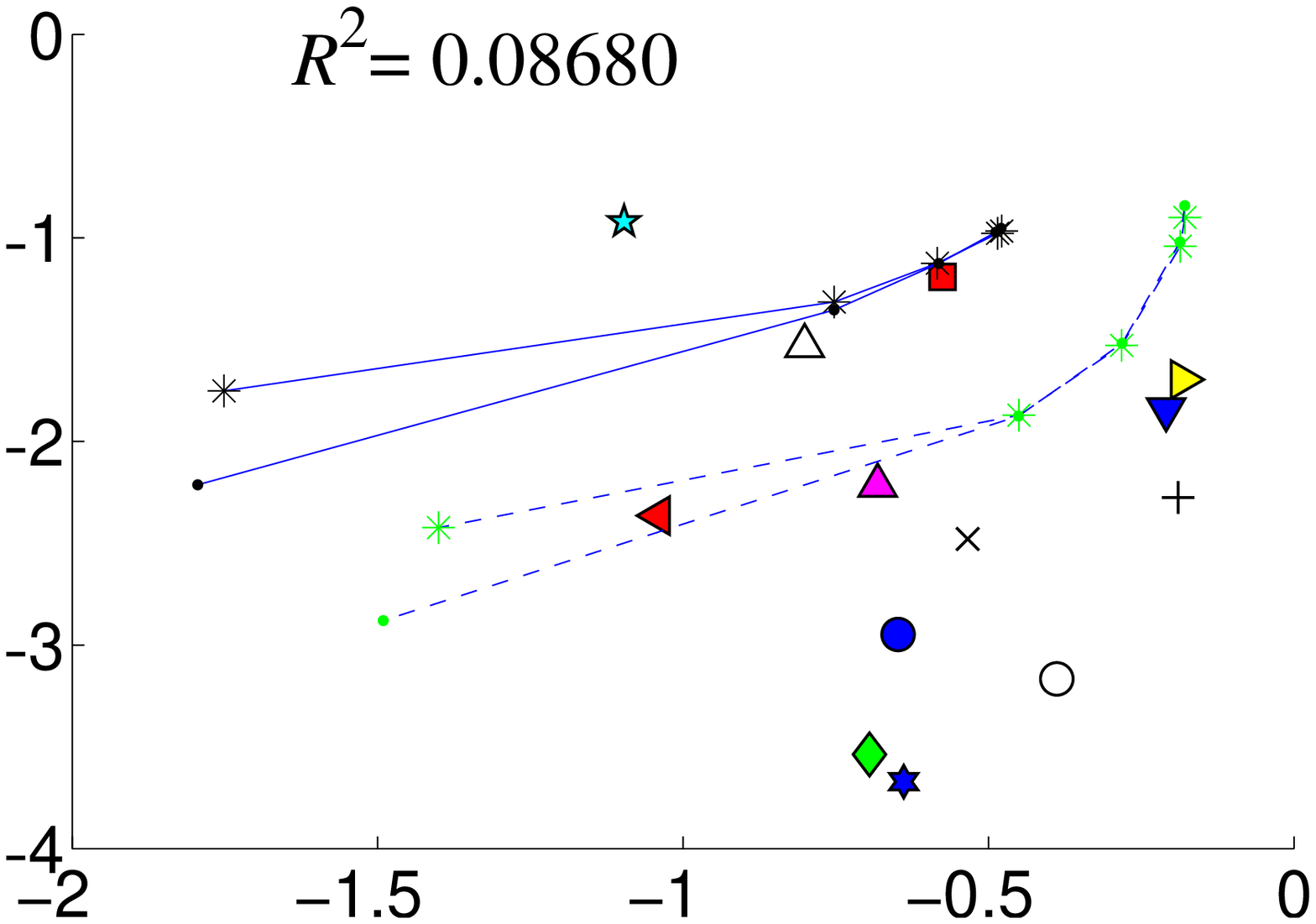} \putpan{\bf(o)} \putx{\bf$\log_{10}(C)$}
\put(-495,0){\begin{rotate}{90}\hspace{11.5cm}\begin{LARGE}$\log_{10} E$\end{LARGE}\end{rotate}}
\end{flushright}
\end{figure*}

\bibliography{networks}
\end{document}